\documentclass{article}
\usepackage{hyperref}
\hypersetup{
colorlinks = true, % false: boxed links; true: colored links
linkcolor=black, % color of internal links
citecolor=black, % color of links to bibliography
urlcolor=black % color of external links
}
\usepackage{amsmath,amsfonts}
\usepackage{amssymb}
\usepackage{booktabs}
\usepackage{graphicx}
\usepackage{paralist}
\usepackage{cleveref}
\usepackage[inkscapelatex=false]{svg}
\usepackage{subcaption}
\usepackage[most]{tcolorbox}
\usepackage{multirow}
\usepackage{xspace}
\usepackage[table,xcdraw]{xcolor}

\newcommand{\inlineimage}[1]{%
\raisebox{-.2\height}{\includegraphics[width=1em]{#1}}%
}
\newcommand{\inlineimageb}[1]{%
\raisebox{-.3\height}{\includegraphics[width=0.7em]{#1}}%
}

\newcommand{\method}{\ensuremath{\mathtt{SB}\text{-}\mathtt{QOPS}}\xspace}

\newcommand{\pauliString}{\ensuremath{\mathit{S}}\xspace}
\newcommand{\pauliFamily}{\ensuremath{\mathit{F}}\xspace}
\newcommand{\numPauliStrings}{\ensuremath{\mathit{|S|}}\xspace}
\newcommand{\eigenvalues}{\ensuremath{\mathit{EV}}\xspace}
\newcommand{\specifiedOutcome}{\ensuremath{\mathit{M}}\xspace}

\newcommand{\expExpValue}{\ensuremath{\mathit{Exp}}\xspace}
\newcommand{\testCase}{\ensuremath{\mathit{T}}\xspace}

\newcommand{\ourPS}{\ensuremath{\mathit{PS_{compact}}}\xspace}
\newcommand{\cut}{CUT\xspace}
\newcommand{\GA}{GA\xspace}
\newcommand{\HC}{HC\xspace}
\newcommand{\RS}{RS\xspace}
\newcommand{\oneplusone}{(1+1) EA\xspace}

\usepackage{PRIMEarxiv}

\title{Search-Based Quantum Program Testing via Commuting Pauli String}

\author{
    Asmar~Muqeet\\
    Simula Research Laboratory \\
    University of Oslo \\
    Oslo\\
    \texttt{asmar@simula.no} \\
    \And
    Shaukat~Ali \\
    Simula Research Laboratory and \\
    Oslo Metropolitan University \\
    Oslo\\
    \texttt{shaukat@simula.no} \\
    \And
    Paolo~Arcaini \\
    National Institute of Informatics \\
    Tokyo\\
    \texttt{arcaini@nii.ac.jp}
}

\begin{document}
\maketitle

\begin{abstract}
Quantum software testing is important for reliable quantum software engineering. Despite recent advances, existing quantum software testing approaches rely on simple test inputs and statistical oracles, costly program specifications, and limited validation on real quantum computers. To address these challenges, we propose \method, a search-based quantum program testing approach via commuting Pauli strings. \method, as a direct extension to a previously proposed QOPS approach, redefines test cases in terms of Pauli strings and introduces a measurement-centric oracle that exploits their commutation properties, enabling effective testing of quantum programs while reducing the need for full program specifications. By systematically exploring the search space through an expectation-value–based fitness function, \method improves test budget utilization and increases the likelihood of uncovering subtle faults. We conduct a large-scale empirical evaluation on quantum circuits of up to 29 qubits on real quantum computers and emulators. We assess three search strategies: Genetic Algorithm, Hill Climbing, and the (1+1) Evolutionary Algorithm, and evaluate \method under both simulated and real noisy conditions. 
Experiments span three quantum computing platforms: IBM, IQM, and Quantinuum. Results show that \method significantly outperforms QOPS, achieving a fault-detection score of 100\% for circuits up to 29 qubits, and demonstrating portability across quantum platforms.
\end{abstract}


\section{Introduction}
Quantum Computing (QC) has introduced new programming paradigms and computational capabilities, giving rise to the need for reliable quantum software engineering practices~\cite{qseRoadmapTOSEM2025}. As a result, Quantum Software Testing~\cite{qstSurveyJSEP2023,quantumSTroadmapTOSEM25} has emerged as a critical research area. Unlike classical software testing, testing quantum programs faces fundamental challenges, including the impossibility of directly observing quantum states without inducing state collapse~\cite{Basics}, the inherent nondeterminism of quantum program outputs~\cite{Basics}, and the limited availability and reliability of current quantum hardware~\cite{quantumSTroadmapTOSEM25}. These characteristics render classical testing techniques largely unsuitable and necessitate the development of novel, quantum-aware testing strategies~\cite{quantumSTroadmapTOSEM25}.

Prior work has proposed a variety of approaches for testing quantum programs, including combinatorial testing~\cite{Combinatorial,qucatASE23tool}, fuzzing~\cite{fuzz}, property-based testing~\cite{property,property2,GarciaICSOC2024}, metamorphic testing~\cite{AbreuQSE22}, and mutation testing~\cite{mutation, mutation2,quantumMutantsEMSE2025,FortunatoICSE22}. Despite this progress, several key challenges remain unresolved in QST. These challenges include the reliance on simplistic input states, the absence of robust and scalable test oracles, the difficulty of handling quantum noise, and the requirement for complete program specifications~\cite{quantumSTroadmapTOSEM25}. In particular, most existing testing methods restrict test inputs to computational basis states $|0\rangle$ and $|1\rangle$. While such inputs may suffice for certain applications, they are incompatible for testing programs that handle input states in superposition, such as quantum search and optimization algorithms~\cite{quantumSTroadmapTOSEM25}. Another major limitation of current approaches is their dependence on statistical oracles that require a full mapping from all possible inputs to their expected output distributions. For larger and more complex quantum programs, constructing such a comprehensive program specification is infeasible, severely limiting the scalability of these methods~\cite{quantumSTroadmapTOSEM25}.

To address these limitations, QOPS~\cite{qops} was proposed as a novel testing approach for quantum programs. Instead of defining test cases through explicit input state initialization, QOPS represents test cases using Pauli strings (i.e., tensor products of Pauli operators), shifting the focus from input initialization to measurement operations for testing. This design makes QOPS particularly suitable for testing programs that inherently rely on superposition, such as search and optimization circuits. Furthermore, QOPS introduces a new oracle that exploits the commuting properties of Pauli strings, which significantly reduces the need for exhaustive program specifications. QOPS is also compatible with modern error mitigation techniques, including the built-in mechanisms provided by IBM’s Estimator API, allowing effective testing of quantum programs on real noisy quantum computers. Despite these advantages, QOPS relies on random search for test case generation, which can be inefficient in terms of both test budget utilization and fault detection effectiveness. Random exploration often fails to optimize the search space, leading to redundant evaluations and limited ability to uncover subtle faults~\cite{ssbse_survey}. In addition, the empirical evaluation of QOPS is restricted in scope, i.e., experiments are conducted on circuits with relatively small sizes (up to 10 qubits) and are limited to the IBM quantum computer.

In this work, we extend the original QOPS approach into \method by introducing a search-based test generation strategy that systematically explores the search space. Instead of relying on unguided random exploration, \method employs an expectation-value–based fitness function that effectively steers the search toward failing test cases. In addition, we improve the empirical evaluation by scaling the analysis to larger quantum circuits of up to 29 qubits.
%To the best of our knowledge, this makes our study one of the few that evaluate a quantum testing approach at such scale, particularly on real quantum computers. 
We further investigate the applicability of \method across multiple quantum computing architectures by conducting experiments on IBM, IQM, and Quantinuum platforms. Finally, we explicitly analyze the impact of quantum noise and study the role of error mitigation techniques in enabling reliable quantum program testing on real quantum computers.

The main contributions of this study are summarized as follows:
\begin{enumerate}
\item A search-based test generation approach combined with an expectation-value–driven fitness function for systematic and efficient exploration of test search space.
\item A large-scale empirical evaluation on circuits with up to 29 qubits, including an analysis of three search strategies: Genetic Algorithm (\GA), Hill Climbing (\HC), and the (1+1) Evolutionary Algorithm (\oneplusone).
\item An extensive assessment of \method under noisy conditions, covering both simulated noise and executions on real quantum computers, with and without error mitigation.
\item A cross-platform evaluation demonstrating the applicability of \method on diverse quantum computing architectures, including IBM, IQM, and Quantinuum.
\end{enumerate}

Our experimental results show that, compared to the original QOPS approach, \method is substantially more effective at uncovering subtle faults. Among the evaluated search strategies, \GA consistently produces higher-quality solutions in terms of fitness optimization, whereas the \oneplusone identifies failing test cases more quickly. Both \GA and \oneplusone detect faults in larger quantum circuits with an average fault detection score of 100\%. Moreover, \method is architecturally portable and can be executed across different quantum computing platforms. However, its effectiveness on real quantum hardware critically depends on the availability of strong error mitigation techniques, such as IBM’s zero-noise extrapolation enhanced with probabilistic error amplification, to ensure reliable test assessment.

\section{Background}\label{sec:background}

\subsection{Quantum Computing}\label{subsec:QC}
This section outlines the basics of quantum computing (QC) and introduces the necessary concepts to understand the rest of the paper.

\paragraph{\textbf{Qubits:}} The {\it quantum bit}, or {\it qubit}, constitutes the fundamental unit of information in quantum computing. Unlike classical bits, which are restricted to binary values of 0 or 1, a qubit can exist in a superposition of the computational basis states $|0\rangle$ and $|1\rangle$. This superposition is described by complex probability amplitudes $(\alpha)$ that encode both magnitude and phase. Using the Dirac notation~\cite{dirac}, the state of a qubit is expressed as a linear combination of basis states, $|\psi\rangle = \alpha_0 |0\rangle + \alpha_1 |1\rangle$. The squared modulus of each amplitude determines the probability of obtaining the corresponding measurement outcome, and these probabilities are constrained by the normalization condition $|\alpha_0|^2 + |\alpha_1|^2 = 1$. 
%In contrast to quantum computation, classical information processing relies on bits that can assume only one of two definite states at any given time.

\begin{figure}[!tb]
\centering
\includegraphics[width=0.99\textwidth]{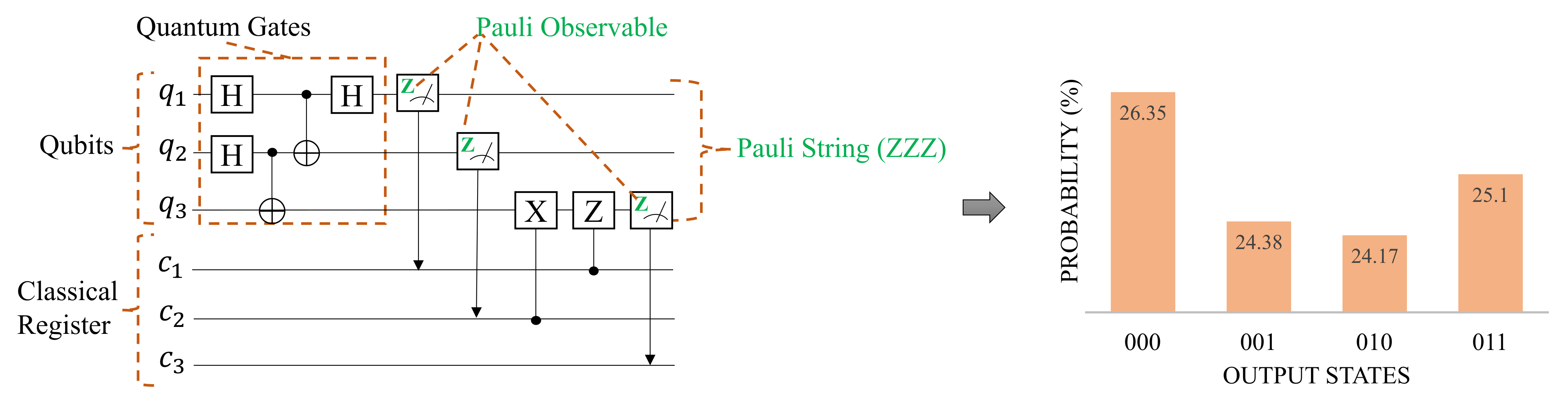}
\caption{A three-qubit quantum teleportation circuit that transfers the quantum state of one qubit to another using entanglement and classical communication.}
\label{fig:teleportation}
\end{figure}

\paragraph{\textbf{Quantum Circuits and Gates:}} Gate-based quantum computers are programmed using {\it quantum circuits}, which consist of sequences of {\it quantum gates} that manipulate the quantum states of qubits. Quantum gates are unitary operators that change a qubit's state based on a unitary matrix~\cite{basic}. A simple example of a quantum gate is the \textit{NOT} gate, which flips a qubit’s state from ($|0\rangle$) to ($|1\rangle$) or vice versa. For better readability, we will refer to a quantum program as a ''circuit'' throughout this paper. In quantum circuits, different types of gates are applied to create {\it superposition} and {\it entanglement}, both of which are fundamental features for quantum information processing. Entanglement is a uniquely quantum phenomenon in which the state of one qubit becomes intrinsically correlated with the state of another, regardless of the physical distance between them. \Cref{fig:teleportation} presents a three-qubit quantum teleportation circuit and its corresponding output.
This circuit implements the quantum teleportation protocol using three qubits. Initially, all qubits ($q_1$, $q_2$, and $q_3$) are in the state $|0\rangle$. The qubit $q_1$ represents the source qubit whose quantum state is to be teleported, while qubits $q_2$ and $q_3$ are used as auxiliary qubits. First, an entangled Bell pair is created between $q_2$ and $q_3$. A \textit{Hadamard} gate~\inlineimage{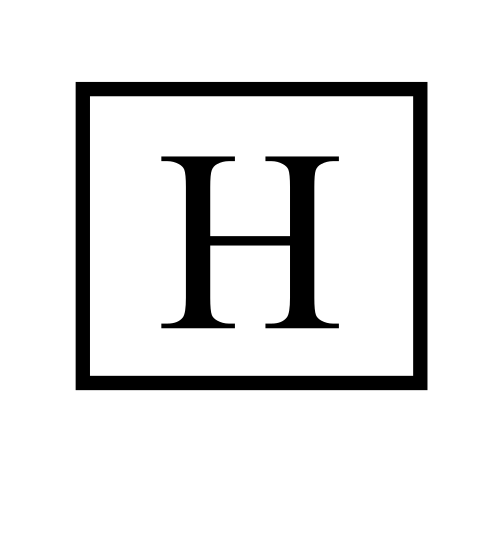} is applied to $q_2$, followed by a \textit{controlled-NOT} gate~\inlineimageb{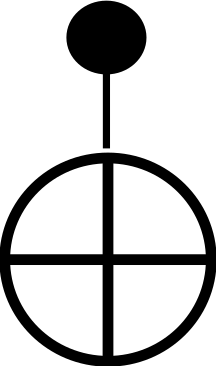} with $q_2$ as the control and $q_3$ as the target. This step establishes entanglement between the two qubits. Next, the source qubit $q_1$ is entangled with $q_2$ by applying a \textit{controlled-NOT} gate with $q_1$ as the control and $q_2$ as the target, followed by a \textit{Hadamard} gate on $q_1$. These operations prepare qubits $q_1$ and $q_2$ for measurement.

A measurement operation~\inlineimage{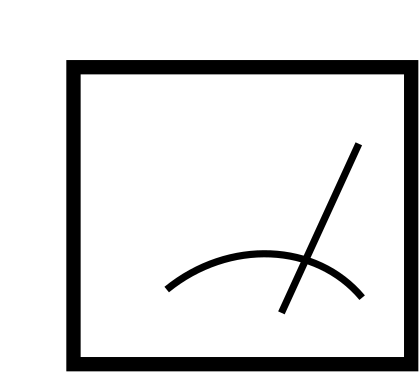} is then performed on qubits $q_1$ and $q_2$, collapsing their joint quantum state into a definite classical outcome. The resulting measurement values are used to classically control conditional Pauli corrections on qubit $q_3$. Specifically, an $X$ gate is applied to $q_3$ if the measurement outcome of $q_2$ is $1$, and a $Z$ gate is applied if the measurement outcome of $q_1$ is $1$. The quantum circuit is executed multiple times to generate a probability distribution of the measurement results. The output consists of binary strings corresponding to the measured qubits, and their associated probabilities reflect how frequently each outcome occurs over repeated executions. After the conditional corrections, qubit $q_3$ holds the same quantum state as the original source qubit $q_1$, i.e., the left- and bottom-most qubit in the output states is a 0 for all possible outcomes. This shows we have a 100\% chance of measuring $q_2$ in state $|0\rangle$, completing the teleportation process.

\paragraph{\textbf{Measurement and Observable:}} Measurement is the mechanism through which the output of a quantum circuit is obtained. The specific form of a measurement depends on the selected \emph{observable}. Observables represent measurable physical quantities, such as a particle’s position, momentum, or spin~\cite{Basics}. Within QC, observables are modeled as Hermitian matrices with fixed eigenvalues~\cite{Basics}, which define the possible outcomes produced by a measurement operation.

In practice, quantum hardware supports four fundamental single-qubit observables ($X$, $Y$, $Z$, and $I$) collectively referred to as \emph{Pauli observables}~\cite{Basics}. When measuring multi-qubit circuits, observables are constructed as tensor products of single-qubit Pauli observables, commonly known as \emph{Pauli strings}. As an illustration, Figure~\ref{fig:teleportation} shows a $Z$ measurement (\inlineimage{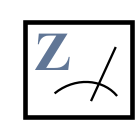}) applied to individual qubits, representing a single-qubit Pauli observable. The complete measurement operation in this circuit corresponds to the Pauli string \emph{ZZZ}, which applies the Pauli $Z$ observable to all qubits $q_1$, $q_2$, and $q_3$.

For a quantum circuit operating on $n$ qubits, the total number of distinct Pauli strings is $4^n$. This exponential growth arises from the four available Pauli observables ($X$, $Y$, $Z$, and $I$) that can be independently assigned to each qubit.

Each Pauli string is associated with a set of \emph{eigenvalues}, which determine the possible outcomes of measuring that string~\cite{basic}. Since Pauli strings are formed by tensor products of Pauli observables (and each Pauli observable is a Hermitian matrix with known eigenvalues), the eigenvalues of a Pauli string are obtained by taking the tensor product of the eigenvalues of its constituent observables. For example, the Pauli observable $Z$ has eigenvalues $(1, -1)$. Consequently, the Pauli string \emph{ZZ} has eigenvalues given by $(1, -1) \otimes (1, -1)$, resulting in the set $(1, -1, -1, 1)$. These eigenvalues are essential for computing expectation values, which are discussed next.

\paragraph{\textbf{Expectation Value:}} Because quantum measurements are inherently probabilistic, repeated executions of the same quantum circuit may yield different outcomes~\cite{basic}. The \emph{expectation value} captures the average result expected from a large number of such measurements and is defined as 
\begin{equation}\label{eg:exp}
E = \sum_{i=1}^{n} e_i P_i,
\end{equation} 
where $n$ denotes the number of possible output states, $P_i$ is the probability of observing the $i$-th outcome, and $e_i$ is the corresponding eigenvalue associated with the Pauli string.

As an example, consider the three-qubit circuit depicted in Figure~\ref{fig:teleportation}, measured using the Pauli string \emph{ZZZ}. The eigenvalues of \emph{ZZZ} are $(1, -1, -1, 1)$. Based on the observed measurement probabilities shown in Figure~\ref{fig:teleportation}, the expectation value is computed as $E = (1 \times 0.2635) + (-1 \times 0.2438) + (-1 \times 0.2417) + (1 \times 0.251) = 0.029$.

\subsection{Commuting Pauli Family}\label{sec:paulifam}
A \emph{Pauli family} is defined as a collection of Pauli strings that satisfy a shared property, such as commutativity~\cite{fastpart}, which is the focus of this work. Two Pauli strings are said to commute if their multiplication is independent of order. For instance, the Pauli strings \emph{ZZ} and \emph{ZI} commute because $ZZ \cdot ZI = ZI \cdot ZZ$.

As discussed in Section~\ref{subsec:QC}, an $n$-qubit quantum circuit admits $4^n$ distinct Pauli strings. These strings can be grouped into $K$ sets of mutually commuting Pauli families, where the value of $K$ depends on the structure of the circuit. Multiple approaches have been proposed to construct such families, with the current state-of-the-art technique introduced by Reggio et al.~\cite{fastpart}.

Grouping Pauli strings into commuting families offers a significant practical benefit: it enables concurrent measurement. In particular, the measurement outcomes obtained from a single Pauli string can be reused to derive expectation values for all other Pauli strings within the same commuting family~\cite{fastpart, Paulistrings2}. This property is widely leveraged in modern quantum computing, especially in applications such as quantum error correction and the acceleration of quantum algorithms~\cite{Paulistrings2}.

\subsection{Quantum Noise}
Quantum computing devices are highly sensitive to quantum noise, which adversely affects the accuracy of their computations. Quantum noise originates from multiple sources. One major source is environmental interference, including magnetic fields and background radiation, which can disrupt quantum operations~\cite{noise_benchmark1}. Such interactions between qubits and their surrounding environment can cause disturbances in quantum states and result in information loss, a process commonly referred to as \emph{decoherence}~\cite{decoherence_def}. 

Another source of noise arises from unintended interactions among qubits themselves. Even when qubits are well isolated from external environments, residual coupling between them can introduce \emph{crosstalk noise}~\cite{crosstalkgatenoise}, leading to unintended correlations and erroneous quantum states. In addition, imperfections in hardware calibration during quantum gate execution further contribute to noise. Small calibration inaccuracies may induce slight deviations in a qubit’s phase or amplitude; while these deviations may be negligible for a single operation, they can accumulate over multiple gates and significantly distort the final quantum state~\cite{crosstalkgatenoise}. 

\begin{figure}[!tb]
\centering
\includegraphics[width=0.9\textwidth]{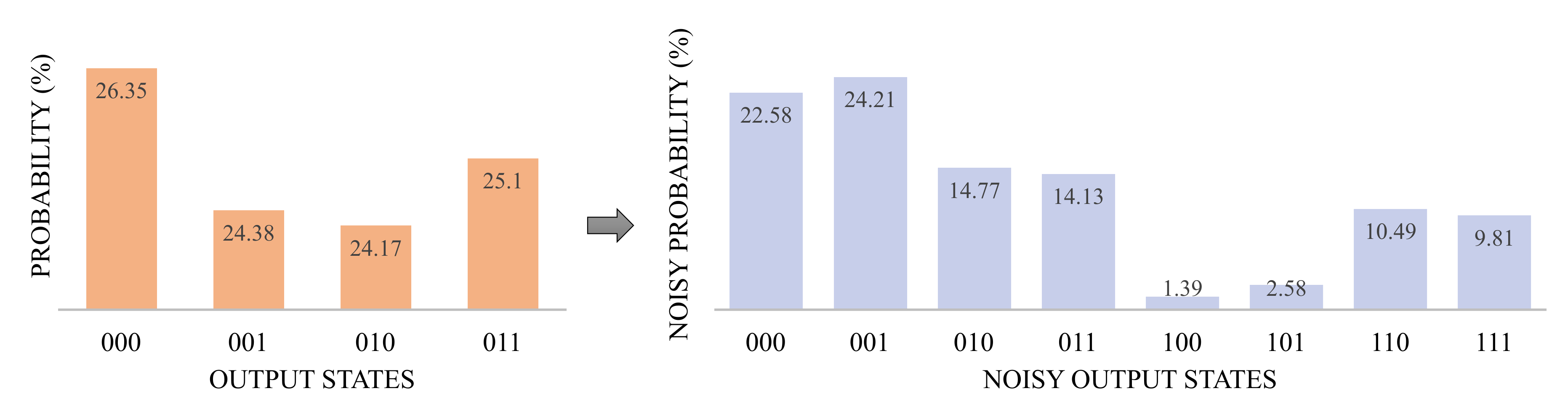}
\caption{Comparison of ideal and noisy outputs for a three-qubit teleportation circuit.}
\label{fig:idealvsnoise}
\end{figure}

It is important to note that noise can affect any qubit at any point during circuit execution, resulting in compounded errors that influence the overall circuit output. \Cref{fig:idealvsnoise} highlights this effect by comparing the ideal and noisy outputs of the three-qubit teleportation circuit shown in \Cref{fig:teleportation}.
In the presence of noise, additional output states such as \texttt{100}, \texttt{101}, \texttt{110}, and \texttt{111} appear, altering the probability distribution of the qubits relative to the ideal case.

Because real quantum hardware is inherently noisy, researchers often employ quantum simulators to execute circuits in noise-free environments~\cite{simulators}. However, ideal quantum simulation is constrained by classical computational resources and becomes infeasible as circuit size grows, even on high-performance computing systems~\cite{simulationlimit}. Consequently, such simulators are typically restricted to circuits involving only a small number of qubits.

To bridge this gap, quantum computing platforms such as IBM and Quantinuum provide configurable \emph{noise models} that approximate the behavior of their physical devices. These noise models emulate effects including decoherence, gate imperfections, and crosstalk, enabling researchers to study the impact of noise and develop error mitigation techniques. Despite their practical value, noisy simulations suffer from the same scalability limitations as ideal simulators and, therefore, remain applicable primarily to small-scale quantum circuits.

\section{Approach}\label{sec:approach}
QOPS~\cite{qops} is a quantum circuit testing approach introduced to address several practical limitations of state-of-the-art (SOTA) quantum circuit testing approaches. A key limitation of existing SOTA approaches is their reliance on overly simplistic test inputs, which are typically restricted to computational basis states $|0\rangle$ and $|1\rangle$. Although such inputs may be adequate for some quantum circuits, they are ill-suited for testing circuits that operate on superposition states, including quantum search and optimization algorithms. In addition, many prior approaches require complete program specifications to serve as test oracles and exhibit limited compatibility with error mitigation methods~\cite{qops}. In contrast to existing approaches, QOPS defines test cases in terms of Pauli strings, eliminating the need for explicit input states and enabling seamless application to quantum search and optimization algorithms. Moreover, QOPS introduces a test oracle that is compatible with commonly adopted error mitigation techniques, facilitating its practical use. By exploiting the commuting properties of Pauli strings, QOPS further reduces the cost associated with full program specification, making quantum circuit testing more scalable and practical than prior SOTA approaches.

Despite these advantages, QOPS relies on random test generation. While this strategy has been shown to outperform existing testing methods in terms of fault detection~\cite{qops}, it is inefficient in terms of resource usage and may fail to effectively expose subtle faults. To address these limitations, we extend the original QOPS approach by incorporating a search-based test generation strategy that systematically guides the exploration of test search space. We refer to the resulting approach as \method: \textbf{S}earch-\textbf{B}ased \textbf{Q}uantum pr\textbf{O}gram testing via \textbf{P}auli \textbf{S}trings. The proposed method requires two inputs from the tester: the \textit{circuit under test} (\cut) and a \textit{compact program specification} (\ourPS). Within \ourPS, the tester specifies the expected measurement outcome of a single Pauli string chosen from a commuting Pauli family, thus significantly reducing the specification cost of full program specification compared to other quantum circuit testing techniques, which typically require complete output specifications~\cite{quitoASE21tool, search, Combinatorial, qlear, qoin}.

To give an example, consider a circuit operating on five qubits. Such a circuit has $4^5 = 1024$ distinct Pauli strings. These strings can be grouped into 33 commuting Pauli families, implying that \ourPS requires at most one Pauli string per family. Consequently, no more than 33 Pauli strings need to be specified, representing a significant reduction relative to the full set of 1024 possibilities.

In the next subsection, we provide the definition of our test cases and the test oracle. The overall working of the approach is described in subsection~\ref{sec:overview}.

\subsection{Test Case and Oracle}~\label{sec:testcase}
\method defines test cases in terms of Pauli strings applied at the measurement stage, rather than by explicitly initializing qubit states as in previous testing approaches~\cite{quitoASE21tool, search, Huang2019, property, property2}. Prior work has shown that combinations of Pauli strings are capable of revealing the same circuit faults that would otherwise be detected through explicit state preparation~\cite{qops}. Traditional test cases rely on configuring the circuit’s initial state in order to observe a particular output behavior. In contrast, equivalent output information can be obtained by modifying the measurement operator through different Pauli strings~\cite{qops}. An important benefit of this measurement-centric testing strategy is its independence from the circuit’s initialization logic, which makes Pauli-string–based test cases particularly well suited for search-based and optimization-oriented quantum circuits

\textbf{Test case definition:} In \method, a test case is constructed as a weighted combination of Pauli strings drawn from a single Pauli family. Formally, a test case is defined as
\begin{equation*}
\testCase = \sum_{i=1}^{\numPauliStrings} c_i \pauliString_i ,
\end{equation*}
where $S$ denotes a selected subset of Pauli strings within the family, $\pauliString_i$ represents the $i$th Pauli string in $S$, and $c_i$ is the corresponding coefficient that determines its contribution.

As an illustrative example, let $S = \{\textit{IZ}, \textit{ZI}, \textit{ZZ}\}$. The resulting test case can then be expressed as
\begin{equation*}
\testCase = c_1 \textit{IZ} + c_2 \textit{ZI} + c_3 \textit{ZZ},
\end{equation*}
where the coefficients $c_1$, $c_2$, and $c_3$ quantify the relative influence of each Pauli string on the aggregated measurement outcome.

\textbf{Test Oracle:} With the proposed test case formulation, a large number of test cases can be constructed by selecting different subsets of Pauli strings. To evaluate whether a circuit under test (\cut) satisfies a given test case, an oracle is required. We introduce the oracle \textit{Expected Expectation} (\expExpValue), which represents the aggregate expectation value produced by executing a test case \testCase on the circuit.

The \expExpValue is defined as the weighted sum of the individual expectation values associated with the Pauli strings comprising the test case:
\begin{equation*}
\expExpValue = \sum_{i=1}^{\numPauliStrings} c_i E_i ,
\end{equation*}
where \numPauliStrings denotes the number of Pauli strings in \testCase, $E_i$ is the expectation value of the $i$th Pauli string, and $c_i$ is its corresponding coefficient.

By applying the expectation value formulation as described in EQ~\ref{eg:exp}, each $E_i$ can be further expanded as a sum over measurement outcomes, yielding
\begin{equation}\label{eq:exp1}
\expExpValue
= \sum_{i=1}^{\numPauliStrings} c_i E_i
= \sum_{i=1}^{\numPauliStrings} c_i \sum_{j=1}^{|\pauliString_i|} w_j P_j ,
\end{equation}
where $\pauliString_i$ refers to the $i$th Pauli string in the test case, $P_j$ is the probability of observing the $j$th output state associated with that Pauli string, and $w_j$ is the corresponding eigenvalue.

Because all Pauli strings within a test case commute, the measurement outcomes obtained from a single Pauli string execution can be reused to compute the expectation values of the remaining strings. Exploiting this property allows Equation~\ref{eq:exp1} to be rewritten in a more compact form:
\begin{equation}\label{eq:exp2}
\expExpValue = \sum_{(i,j)\in I} c_i w_{i,j} M_j ,
\end{equation}
where the index set is defined as $I = \{(i,j) \mid 1 \leq i \leq \numPauliStrings,; 1 \leq j \leq \pauliString_i\}$. Here, $e_{i,j}$ denotes the $j$th eigenvalue of the $i$th Pauli string, $M_j$ represents the observed value of the $j$th measurement outcome as specified in \ourPS, and $c_i$ is the weight assigned to the $i$th Pauli string in the test case.

\subsection{Overview}\label{sec:overview}
\begin{figure}[!tb]
\centering
\includegraphics[width=1\textwidth]{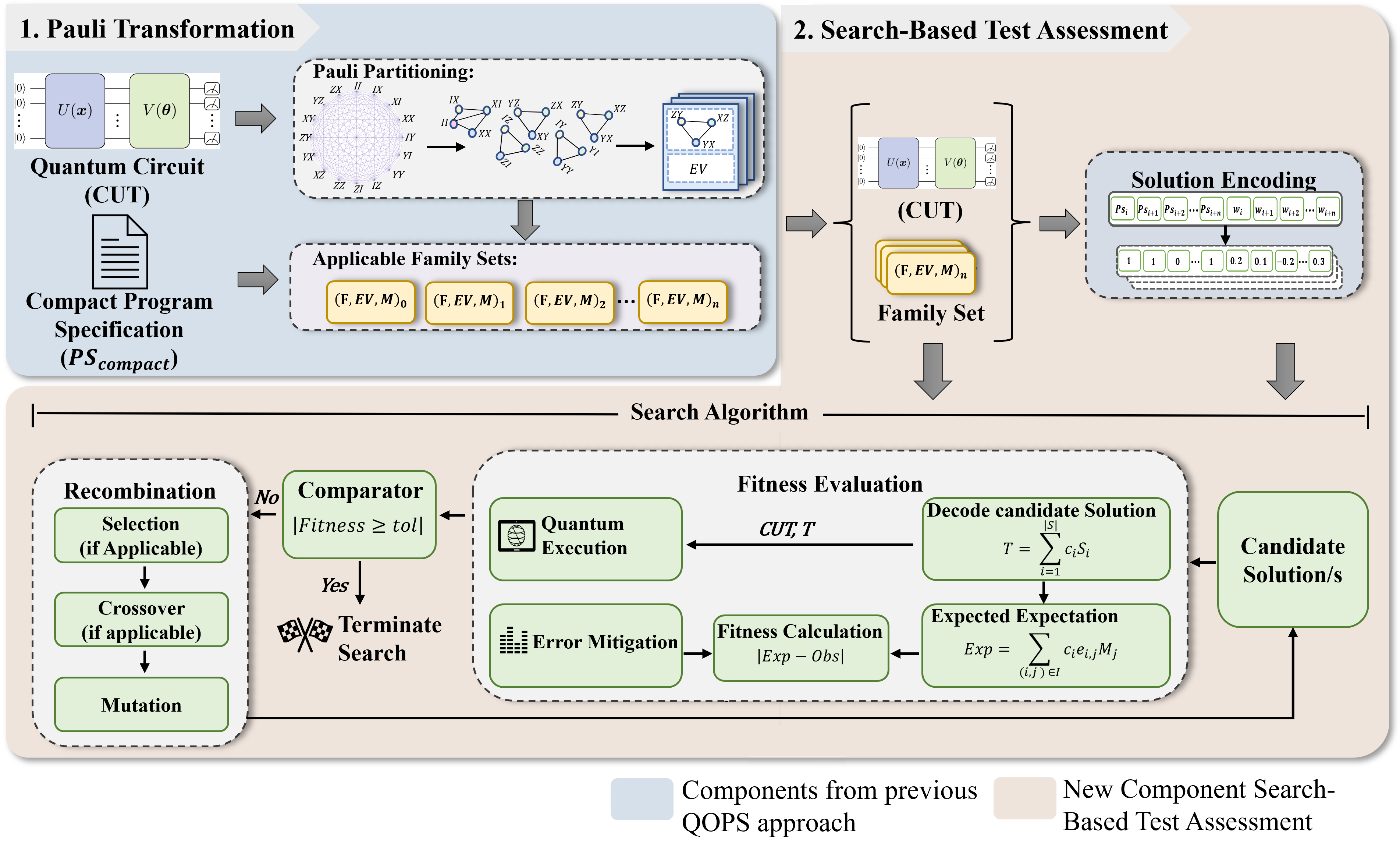}
\caption{Overview of \method: The set $(\pauliFamily, \eigenvalues, \specifiedOutcome)$ represents the core elements of \method. It consists of a Pauli family $\pauliFamily$, the corresponding eigenvalues $\eigenvalues$ associated with the Pauli strings in that family, and a specified measurement outcome $\specifiedOutcome$. The specified outcome refers to a single Pauli string selected from $\pauliFamily$ and is determined according to a compact program specification, denoted as \ourPS.}
\label{fig:overview}
\end{figure}

Figure~\ref{fig:overview} presents an overview of the \method approach, which is composed of two primary components.
The first component, \textit{Pauli Transformation}, follows the same procedure introduced in earlier work~\cite{qops}. The second component, \textit{Search-Based Test Assessment}, is newly introduced in this study. Details of the components are described in the subsequent subsections.

\subsubsection{Pauli Transformation}
The \method workflow begins by constructing families of commuting Pauli strings for testing purposes, using the \cut and the \ourPS as inputs. To identify these families, \method adopts the Pauli partitioning algorithm proposed by Reggio et al.~\cite{fastpart}. While a detailed description of the algorithm is available in the original work, its operation can be summarized here at a high level. The partitioning process has two stages. First, the algorithm identifies subsets of Pauli strings that satisfy mutual commutativity. For each such subset and a given \cut, it then attempts to compute a transformation matrix that simultaneously diagonalizes all Pauli strings in the subset. If such a matrix exists, the subset is accepted as a valid Pauli family. This procedure is iteratively applied until no additional commuting sets can be formed.

The output of this stage is a collection of $K$ Pauli families, along with the eigenvalues corresponding to each Pauli string within a family. Given the measurement outcome of any Pauli string within a family, these eigenvalues enable the computation of the test oracle for test cases constructed from arbitrary subsets of strings in that family~\cite{qops}.

Following the identification of Pauli families and their eigenvalues, \method evaluates the \ourPS to discard families that are not applicable for testing. Because \ourPS specifies expected outcomes for only a limited number of Pauli strings, not all identified families may be usable. In the ideal case, the specification includes at least one Pauli string per family, allowing all families to participate in the testing process. However, when fewer Pauli strings are specified, the number of applicable families is reduced accordingly.

As illustrated in Figure~\ref{fig:overview}, the Pauli Transformation stage ultimately produces a filtered set of Pauli families. Each family in this set is represented as a set comprising three elements: \textbf{\pauliFamily}, which identifies the commuting Pauli family; \textbf{\eigenvalues}, which stores the eigenvalues associated with the Pauli strings in that family; and \textbf{\specifiedOutcome}, which denotes the measurement outcome provided in \ourPS for one Pauli string selected from \textbf{\pauliFamily}.

\subsubsection{Search-Based Test Assessment}
The search-based test assessment phase takes as input the circuit under test (\cut) together with the set of Pauli families produced by the Pauli Transformation stage, represented as $\{\textbf{\pauliFamily}, \textbf{\eigenvalues}, \textbf{\specifiedOutcome}\}$. The workflow begins by constructing a solution encoding, which defines both the search space and the structure used to encode candidate solutions.

Based on the solution encoding and the chosen search strategy, an initial population of candidate solutions is generated either through random initialization or other initialization methods, such as heuristic approaches, and evaluated using a predefined fitness function. If a satisfactory solution is identified during this evaluation phase, the search process terminates. Otherwise, the strategy proceeds to a recombination stage. Depending on the employed search strategy, this stage may include operations such as selection, crossover, and mutation to generate new candidate solutions. These newly produced candidates are added to the population and subsequently evaluated.

The search–evaluate–recombine cycle continues iteratively until either a solution is found or a user-defined computational budget is exhausted. Each stage of this workflow is described in detail in the following subsections.

\begin{figure}[!tb]
\centering
\includegraphics[width=0.8\textwidth]{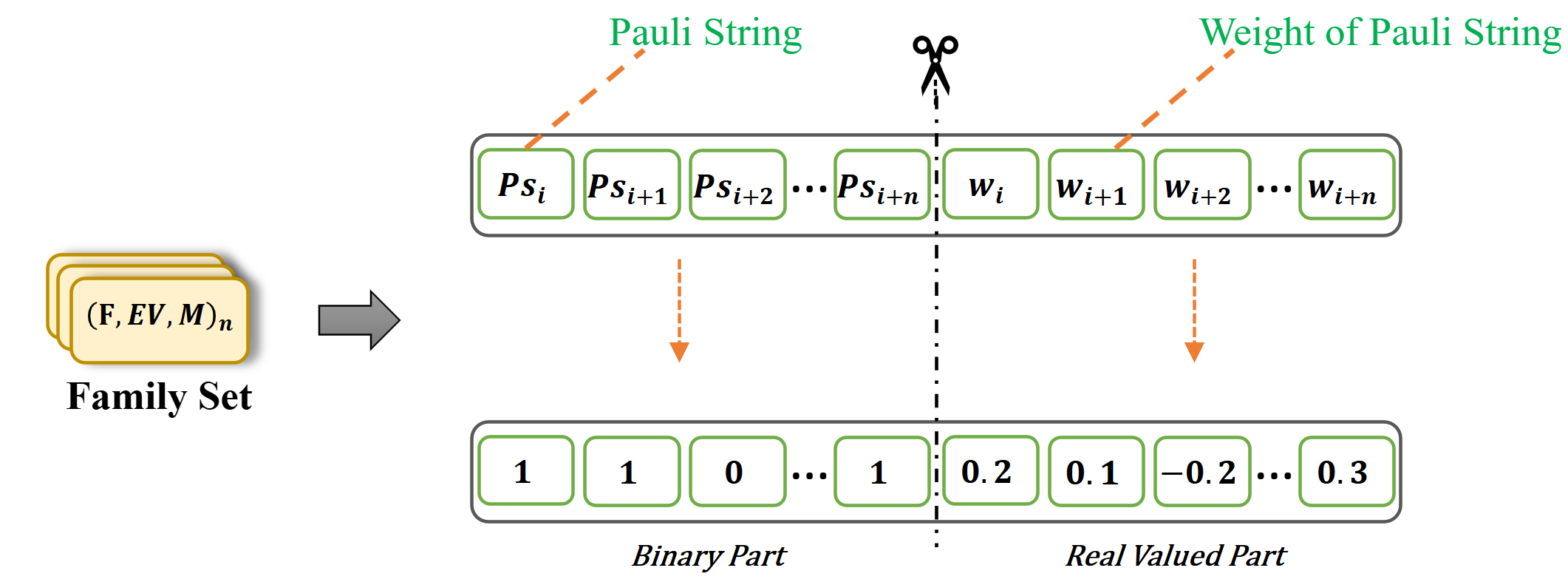}
\caption{Linear array–based solution encoding used to represent candidate solutions in the search strategy}
\label{fig:candidate}
\end{figure}

\paragraph{\bf Solution Encoding}
This component takes the set of Pauli families $\{\textbf{\pauliFamily}, \textbf{\eigenvalues}, \textbf{\specifiedOutcome}\}$ as input and constructs a solution encoding for each family. To ensure broad applicability across different search strategies, we adopt a simple and uniform candidate representation based on a linear array. The number of Pauli strings available within the selected Pauli family determines the length of the linear array. Figure~\ref{fig:candidate} shows the linear array that encodes the candidate solution.
The array is divided into two segments. The first segment uses a binary encoding to indicate whether a given Pauli string is included in the test case, while the second segment consists of real-valued entries that encode the corresponding weights assigned to the selected Pauli strings.

For a given Pauli family \textbf{\pauliFamily}, the size of the candidate array is defined as
\begin{equation*}
\min\bigl(2 \times \text{sizeof}(\textbf{\pauliFamily}), 64\bigr).
\end{equation*}
The upper bound of 64 elements is imposed to constrain the overall size of the search space and, consequently, the computational resources required for exploration. This value was chosen based on preliminary empirical analysis, which indicated that a maximum test case length of 32 Pauli strings was sufficient to expose all observed faults.

\paragraph{\bf Fitness Evaluation} The fitness evaluation component assesses the fitness of a single candidate solution by computing its fitness score. The fitness function is defined as
\begin{equation}\label{eq:fitness}
\lvert \textit{Exp} - \textit{Obs} \rvert,
\end{equation}
where \textit{Exp} denotes the expected expectation value obtained from the oracle described in Section~\ref{sec:testcase}, and \textit{Obs} represents the expectation value observed after executing the corresponding test case on a quantum computer or simulator.

To perform this evaluation, the candidate solution is first decoded into a concrete test case according to the test case formulation in Section~\ref{sec:testcase}. As an illustrative example, consider a two-qubit Pauli-\textit{Z} family defined as $\textbf{\pauliFamily} = \{\textit{II}, \textit{IZ}, \textit{ZI}, \textit{ZZ}\}$ and a candidate solution encoded as
\begin{equation*}
S = [0, 1, 1, 0, 0.1, 0.2, 0.5, 0.2].
\end{equation*}
The binary segment of $S$ indicates that the Pauli strings \textit{IZ} and \textit{ZI} are selected, while the remaining entries specify their corresponding weights. Consequently, the decoded test case $T$ is given by
\begin{equation*}
T = 0.2\textit{IZ} + 0.5\textit{ZI},
\end{equation*}
which follows directly from the test case definition presented in Section~\ref{sec:testcase}.

The first step in the fitness computation is the expected expectation value (\textit{Exp}). To illustrate this process, consider a Pauli family set ${\pauliFamily, \eigenvalues, \specifiedOutcome}$ obtained from the Pauli Transformation stage, where
\begin{equation*}
\pauliFamily = \{\textit{II}, \textit{IZ}, \textit{ZI}, \textit{ZZ}\},
\end{equation*}
and the corresponding eigenvalues are
\begin{equation*}
(1, 1, 1, 1), (1, -1, 1, -1), (1, 1, -1, -1), (1, -1, -1, 1).
\end{equation*}
Assume that the compact program specification \ourPS defines the measurement outcome for the Pauli string \textit{ZZ} as
\begin{equation*}
\specifiedOutcome = \{\textit{ZZ} : \{00: 0,; 01: 0.5,; 10: 0.3,; 11: 0.2\}\}, 
\end{equation*}
where the output states $(00,$ $01,$ $10,$ $11)$ are associated with the corresponding probabilities $(0,$ $0.5,$ $0.3,$ $0.2)$. Because the Pauli strings \textit{ZZ}, \textit{ZI}, and \textit{IZ} belong to the same commuting family, the measurement outcomes obtained for \textit{ZZ} can be reused to compute the expected expectation value of the test case \testCase. Since \testCase includes the Pauli strings \textit{IZ} and \textit{ZI}, their corresponding eigenvalue vectors are $(1, -1, 1, -1)$ and $(1, 1, -1, -1)$, respectively.

By substituting these values into Equation~\ref{eq:exp2}, the expected expectation value is computed as
\begin{equation*}
\expExpValue = (c_0 w_{0,0} M_0 + c_0 w_{0,1} M_1 + c_0 w_{0,2} M_2 + c_0 w_{0,3} M_3)
\end{equation*}
\begin{equation*}
+ (c_1 w_{1,0} M_0 + c_1 w_{1,1} M_1 + c_1 w_{1,2} M_2 + c_1 w_{1,3} M_3),
\end{equation*}
which, after substituting the concrete values, yields
\begin{equation*}
\expExpValue = (0.2 \cdot 1 \cdot 0 + 0.2 \cdot (-1) \cdot 0.5 + 0.2 \cdot 1 \cdot 0.3 + 0.2 \cdot (-1) \cdot 0.2)
\end{equation*}
\begin{equation*}
+ (0.5 \cdot 1 \cdot 0 + 0.5 \cdot 1 \cdot 0.5 + 0.5 \cdot (-1) \cdot 0.3 + 0.5 \cdot (-1) \cdot 0.2) \
= -0.08.
\end{equation*}

Accordingly, the expected expectation value associated with the test case \testCase is $-0.08$.

Once the oracle value \textit{Exp} has been computed, the corresponding test case \testCase is executed on the \cut using either a quantum simulator or a quantum computer. This execution yields the observed expectation value, denoted as \textit{Obs}. For instance, running \testCase on IBM’s real quantum computer without error mitigation produces an observed value of $\textit{Obs} = -0.1$.

The fitness score is then obtained by substituting \textit{Exp} and \textit{Obs} into the fitness function defined in Equation~\ref{eq:fitness}:
\begin{equation*}
\lvert -0.08 - (-0.1) \rvert = 0.02,
\end{equation*}
resulting in a fitness value of $0.02$. The computed fitness score is compared against a user-defined threshold to determine whether the corresponding test case is considered to have failed, in which case the search procedure terminates. If no test case exceeds the specified threshold, the search process continues until the allocated computational budget is exhausted.

\section{Experiment Design}\label{sec:experimentDesign}
To assess the applicability and effectiveness of \method, we address the following research questions:

\begin{itemize}
\item[\textbf{RQ1}] How effective is \method in detecting faults compared to the random-search–based QOPS approach?
\item[\textbf{RQ2}] What is the impact of quantum noise on the fault-detection capability of \method?
\item[\textbf{RQ3}] Is \method applicable and effective when executed on real quantum hardware?
\end{itemize}

\subsection{Benchmarks}
In our previous study~\cite{qops}, we evaluated the approach using the MQT benchmark~\cite{mqt}, which provides a diverse collection of quantum circuits derived from real-world applications. To ensure feasible execution times on both quantum simulators and physical devices, the evaluation was restricted to circuits containing between 2 and 10 qubits. For these circuits, faulty variants were generated using Muskit~\cite{mutation} by applying all mutation operators defined in the tool.

However, the mutants produced by Muskit were relatively easy to detect, as the applied mutations introduced big changes in the output of the circuits. Muskit was primarily used to enable direct comparison with existing testing oracles and methodologies, thereby demonstrating the effectiveness of the QOPS approach. In this context, a random search strategy was sufficient to detect all faults that could be identified by other testing techniques.

In contrast, random search is less suitable when the objective is to identify more subtle faults and to efficiently utilize a limited testing budget. Consequently, in the present evaluation, we selected 10 real-world application circuits from the MQT benchmark and manually constructed three faulty circuits per \cut that are more challenging to detect using existing testing oracles. These faulty circuits were created by carefully analyzing the program logic and introducing rotational gates with very small angles, thereby inducing minor deviations in the output behavior. Such subtle faults are difficult to expose using other test oracles, such as chi-square–based methods~\cite{quitoASE21tool, Combinatorial, search}, making them well suited for evaluating the effectiveness of \method. In addition to the faulty circuits, we also construct an equivalent circuit for each \cut and qubit count. The equivalent circuit corresponds to the original circuit under test without any injected fault and represents the non-faulty class.

To further assess scalability and broaden applicability, we extended the evaluation beyond the original 2–10 qubit range to include circuits with 5, 10, 15, 20, 25, and 29 qubits, enabling a systematic analysis of the testing approach at larger scales. The upper limit of 29 qubits reflects the practical constraints of current quantum simulators, which can simulate circuits of this size with reasonable execution times and maximal GPU utilization~\cite{simulators,simulationlimit}. To the best of our knowledge, this makes our study one of the few evaluations of quantum program testing techniques conducted at such a scale.

Under this configuration, the experimental setup comprises a total of 180 quantum circuits, corresponding to 10 base programs, three manually constructed circuits per \cut, and six distinct qubit counts.

\subsection{Experiment Settings}\label{sec:settings}
The implementation of \method, along with the experimental results, is publicly available in our repository~\cite{sourcecode}.
\subsubsection{Search strategies}
In our implementation, we used three representative search strategies widely adopted in search-based software engineering (SBSE)~\cite{ssbse_survey}. Specifically, we chose Genetic Algorithm (\GA) for its strong global exploration capabilities, Hill Climbing (\HC) as a representative local search strategy, and the 1+1 Evolutionary Algorithm (\oneplusone) for its simplicity and computational efficiency. To implement the search strategies, we use the Mealpy: MEta-heuristic ALgorithms in PYthon library~\cite{mealpy, mealpyoptimizer}, which provides implementations of 140 metaheuristic search strategies. Candidate solutions are encoded using a linear architecture (see Section~\ref{sec:approach}) defined through Mealpy’s \textit{Problem} class, enabling seamless integration with all supported search strategies. While this study evaluates \GA, \HC, and the \oneplusone, the abstract problem formulation allows \method to be readily extended to any of the search strategies available in Mealpy without modification to the candidate representation.

We set the total computational budget for \method to 140 fitness evaluations. Under this budget, the \GA is configured with a population size of 40 over 10 generations, \HC and the \oneplusone are each executed for 140 fitness evaluations. These parameter settings were determined through a preliminary grid search conducted on circuits with five qubits, with the objective of identifying configurations under which all three search strategies achieve their best performance.

\subsubsection{Circuit Execution (ideal)}
We used Qiskit's AER-simulator~\cite{qiskit} to derive the \ourPS required by \method. In contrast to the previous QOPS evaluation, which considered all Pauli families, the present study restricts \ourPS to the Z-Pauli family. This choice is motivated by scalability considerations: to the best of our knowledge, existing Pauli partitioning algorithms are unable to efficiently partition all Pauli families for circuits with up to 29 qubits within a reasonable computation time. Since Pauli partitioning itself is not the focus of this work, we adopt the Z-Pauli family, which is also the default measurement basis for most real quantum computing platforms, including those provided by IBM~\cite{IBMQuantum}, Google~\cite{GoogleQuantum}, IQM~\cite{IQMQuantum}, and Quantinuum~\cite{QuantinuumQuantum}. This design choice enables maximal qubit utilization while remaining aligned with practical hardware constraints.

\method requires a tolerance threshold to determine whether a test case is classified as failing. To ensure a fair comparison with the previous QOPS approach, we adopt the same threshold value used in that study, namely 1\%, for evaluations conducted on the quantum simulator. For experiments on real quantum computers and noisy simulations, the presence of quantum noise makes the selection of a fixed threshold less straightforward. Consequently, we do not impose a predefined threshold in these settings. Instead, we execute the search strategies for the maximum computational budget and subsequently analyze the results under different threshold configurations, informed by the findings from RQ1.

\subsubsection{Noise Simulation}
To study the impact of quantum noise, we employ noise models provided by IBM for three publicly accessible quantum devices (IBM Brisbane, IBM Torino, and IBM Kingston) and execute circuits using IBM’s Estimator API~\cite{qiskit}. These noise models are selected because they are frequently calibrated and updated by IBM to closely reflect the behavior of their corresponding physical devices.

\subsubsection{Real Quantum Computers}
For experiments on real quantum computers, we utilize quantum computers from IBM and IQM, as well as Quantinuum’s quantum simulator, thereby demonstrating the applicability of \method across multiple quantum computing architectures and its effectiveness on both simulated and real quantum platforms.

\subsubsection{Error Mitigation}
With respect to error mitigation on real quantum hardware, the most commonly adopted techniques include probabilistic error cancellation (PEC), zero-noise extrapolation (ZNE), and Clifford data regression (CDR)~\cite{mitiq}. IBM’s Estimator API provides built-in support for error mitigation based on ZNE and PEC. PEC incurs an exponential computational overhead~\cite{peccost}. Combined with the monthly usage limits of IBM quantum devices, this makes PEC impractical for our experiments. Consequently, we employ ZNE as the error mitigation strategy for experiments conducted on IBM devices.

For experiments using noise models, we apply an open-source implementation of ZNE provided by Mitiq~\cite{mitiq}. In contrast, for executions on IQM and Quantinuum systems, no error mitigation techniques are currently available or directly supported. Consequently, all experiments on these platforms are performed without error mitigation.

\subsection{Evaluation Metrics}\label{metrics}
To address RQ1, we evaluate the effectiveness of different search strategies in detecting faults in the circuit under test. For each combination of circuit, qubit count, and search strategy, we perform 10 independent runs to account for the stochastic nature of the search processes. In each run, the search is executed separately for each faulty circuit associated with the \cut for each qubit count.

A fault is considered detected if the search generates a test case that fails within the allocated computational budget. Detection is encoded as a binary outcome: a value of 1 indicates successful detection of the fault, while a value of 0 indicates failure to detect it. For each run, we compute a {\it Fault Detection Score} (FDS) by averaging the detection outcomes across the three faulty circuits of the \cut. Formally, for run $r$, the fault detection score is defined as:
\begin{equation*}
\text{FDS} = \frac{1}{3} \sum_{i=1}^{3} d_i,
\end{equation*}
where $d_i \in {0,1}$ denotes whether fault $i$ was detected during run $r$. To reduce the impact of randomness inherent in the search strategies, we repeat this procedure for 10 runs and compute the final metric as the mean of the fault detection scores across all runs:
\begin{equation*}
\text{AvgFDS} = \frac{1}{10} \sum_{r=1}^{10} \text{FDS} .
\end{equation*}

The resulting \textit{AvgFDS} value lies in the range $[0,1]$ and represents the average proportion of faults detected for a given circuit, qubit count, and search strategy. A value of 0 indicates that no faults were detected in any run, whereas a value of 1 indicates that all faults were consistently detected across all runs. To assess the statistical significance of performance differences among the search strategies, we utilized the Wilcoxon signed-rank test, and Cliff’s delta effect size as per the recommendation in~\cite{recomended_effectsize, statistics2} for pairwise comparisons.

For RQ2 and RQ3, to account for the impact of quantum noise on search-based test assessment, we evaluate the performance of \method under different tolerance threshold configurations using a binary classification framework. In this setting, the ability of an approach to correctly distinguish between faulty circuits and equivalent (i.e., non-faulty) circuits serves as an indicator of its effectiveness in the presence of noise. The test assessment outcome of whether a failing test case is identified for a given threshold is treated as the classification decision.

Using this formulation, we compute commonly used classification quality metrics, including precision, recall, F1-score, as well as the numbers of false positives and false negatives, following their standard definitions~\cite{commonf3}. These metrics enable a detailed analysis of the trade-offs introduced by different threshold values, particularly in noisy execution environments. Similarly to RQ1, we repeat the evaluation process 10 times to mitigate the effects of randomness. In the context of RQ2 and RQ3, these repeated executions account for variability arising from the stochastic nature of the search strategy, quantum noise during circuit execution, and the randomness in the error mitigation method.

\section{Results}\label{sec:experimentalResults}

\subsection{RQ1---Search-Based test assessment}

\subsubsection{Fault Detection Analysis}
To address RQ1, we compare the effectiveness of different search strategies in detecting faults across multiple quantum circuits and qubit counts using the AvgFDS metric defined in Section~\ref{metrics}. As described earlier, AvgFDS represents the average proportion of detected faults over ten runs to account for randomness in search strategies. Table~\ref{tab:RQ1mds} reports the AvgFDS results for the baseline random search used in the previous QOPS approach (\RS), compared with \GA, \HC, and the \oneplusone, evaluated across qubit counts ranging from 5 to 29 and across multiple quantum circuits.
\begin{table}[!tb]
\caption{RQ1 -- Comparison of search strategies for search-based test assessment. The row labeled \textbf{\RS} reports the results of the random search strategy used in the prior QOPS approach~\cite{qops}. For each circuit and qubit count, the corresponding AvgFDS value is reported, with the highest value highlighted in red. The \textbf{Average} column presents the mean AvgFDS computed across all quantum circuits for each qubit count.}
\label{tab:RQ1mds}
\resizebox{1\columnwidth}{!}{%
\begin{tabular}{c|c|cccccccccc|c}
\toprule
& & \multicolumn{10}{c|}{\textbf{Circuits}} & \\
\multirow{-2}{*}{\textbf{Strategies}} & \multirow{-2}{*}{\textbf{Qubits}} & \textbf{Ae} & \textbf{Dj} & \textbf{Ghz} & \textbf{Graph} & \textbf{Qnn} & \textbf{Random} & \textbf{Real} & \textbf{Su2} & \textbf{Twolocal} & \multicolumn{1}{c|}{\textbf{Wstate}} & \multirow{-2}{*}{\textbf{Average}} \\
\midrule
& \textbf{5} & 0.33 & 0.33 & 0.90 & 0.33 & 0.63 & 0.23 & 0.00 & 0.23 & 0.00 & 0.00 & 0.30 \\
& \textbf{10} & 0.37 & 0.33 & \cellcolor[HTML]{FFCCC9}1.00 & 0.00 & 0.53 & 0.07 & 0.10 & 0.03 & 0.07 & 0.10 & 0.26 \\
& \textbf{15} & 0.33 & 0.33 & 0.90 & 0.07 & 0.03 & 0.00 & 0.00 & 0.07 & 0.07 & 0.27 & 0.21 \\
& \textbf{20} & 0.37 & 0.33 & \cellcolor[HTML]{FFCCC9}1.00 & 0.00 & 0.77 & 0.03 & 0.00 & 0.03 & 0.03 & 0.07 & 0.26 \\
& \textbf{25} & 0.33 & 0.33 & \cellcolor[HTML]{FFCCC9}1.00 & 0.03 & 0.00 & 0.07 & 0.03 & 0.10 & 0.07 & 0.03 & 0.20 \\
\multirow{-6}{*}{\textbf{\RS}} & \textbf{29} & 0.47 & 0.37 & 1.00 & 0.10 & 0.07 & 0.00 & 0.03 & 0.00 & 0.10 & 0.03 & 0.22 \\ \hline
& \textbf{5} & 0.50 & 0.70 & 0.97 & \cellcolor[HTML]{FFCCC9}0.93 & \cellcolor[HTML]{FFCCC9}1.00 & \cellcolor[HTML]{FFCCC9}0.83 & 0.67 & \cellcolor[HTML]{FFCCC9}0.93 & \cellcolor[HTML]{FFCCC9}0.80 & 0.07 & 0.74 \\
& \textbf{10} & 0.73 & \cellcolor[HTML]{FFCCC9}1.00 & 0.97 & \cellcolor[HTML]{FFCCC9}1.00 & \cellcolor[HTML]{FFCCC9}1.00 & \cellcolor[HTML]{FFCCC9}1.00 & \cellcolor[HTML]{FFCCC9}1.00 & \cellcolor[HTML]{FFCCC9}1.00 & \cellcolor[HTML]{FFCCC9}1.00 & \cellcolor[HTML]{FFCCC9}1.00 & \cellcolor[HTML]{FFCCC9}0.97 \\
& \textbf{15} & 0.83 & \cellcolor[HTML]{FFCCC9}1.00 & 0.97 & 0.97 & \cellcolor[HTML]{FFCCC9}1.00 & 0.97 & \cellcolor[HTML]{FFCCC9}1.00 & \cellcolor[HTML]{FFCCC9}1.00 & \cellcolor[HTML]{FFCCC9}1.00 & 0.90 & \cellcolor[HTML]{FFCCC9}0.96 \\
& \textbf{20} & \cellcolor[HTML]{FFCCC9}1.00 & 0.83 & \cellcolor[HTML]{FFCCC9}1.00 & \cellcolor[HTML]{FFCCC9}1.00 & \cellcolor[HTML]{FFCCC9}1.00 & 0.97 & \cellcolor[HTML]{FFCCC9}1.00 & \cellcolor[HTML]{FFCCC9}1.00 & 0.97 & \cellcolor[HTML]{FFCCC9}1.00 & 0.98 \\
& \textbf{25} & \cellcolor[HTML]{FFCCC9}1.00 & \cellcolor[HTML]{FFCCC9}1.00 & 0.93 & \cellcolor[HTML]{FFCCC9}1.00 & \cellcolor[HTML]{FFCCC9}1.00 & 0.97 & 0.97 & \cellcolor[HTML]{FFCCC9}1.00 & 0.97 & \cellcolor[HTML]{FFCCC9}1.00 & \cellcolor[HTML]{FFCCC9}0.98 \\
\multirow{-6}{*}{\textbf{\GA}} & \textbf{29} & \cellcolor[HTML]{FFCCC9}1.00 & 0.90 & 0.96 & 0.96 & 0.96 & 0.90 & 0.96 & 0.93 & \cellcolor[HTML]{FFCCC9}1.0 & \cellcolor[HTML]{FFCCC9}1.0 & 0.95 \\ \hline
& \textbf{5} & 0.40 & 0.50 & 0.03 & 0.33 & 0.73 & 0.57 & 0.23 & 0.67 & 0.30 & 0.27 & 0.40 \\
& \textbf{10} & 0.70 & 0.80 & 0.40 & 0.67 & 0.33 & 0.37 & 0.40 & 0.57 & 0.37 & 0.07 & 0.47 \\
& \textbf{15} & 0.80 & \cellcolor[HTML]{FFCCC9}1.00 & 0.63 & 0.30 & 0.33 & 0.40 & 0.47 & 0.37 & 0.37 & 0.10 & 0.48 \\
& \textbf{20} & 0.77 & 0.77 & 0.17 & 0.43 & 0.30 & 0.40 & 0.27 & 0.47 & 0.47 & 0.43 & 0.45 \\
& \textbf{25} & 0.87 & 0.53 & 0.27 & 0.37 & 0.37 & 0.33 & 0.40 & 0.43 & 0.33 & 0.40 & 0.43 \\
\multirow{-6}{*}{\textbf{\HC}} & \textbf{29} & 0.70 & 0.70 & 0.53 & 0.36 & 0.40 & 0.30 & 0.33 & 0.36 & 0.36 & 0.26 & 0.43 \\ \hline
& \textbf{5} & \cellcolor[HTML]{FFCCC9}0.87 & \cellcolor[HTML]{FFCCC9}1.00 & \cellcolor[HTML]{FFCCC9}1.00 & 0.73 & 0.80 & \cellcolor[HTML]{FFCCC9}0.83 & \cellcolor[HTML]{FFCCC9}1.00 & \cellcolor[HTML]{FFCCC9}0.93 & 0.60 & \cellcolor[HTML]{FFCCC9}0.43 & \cellcolor[HTML]{FFCCC9}0.82 \\
& \textbf{10} & \cellcolor[HTML]{FFCCC9}0.93 & 0.73 & \cellcolor[HTML]{FFCCC9}1.00 & \cellcolor[HTML]{FFCCC9}1.00 & \cellcolor[HTML]{FFCCC9}1.00 & \cellcolor[HTML]{FFCCC9}1.00 & \cellcolor[HTML]{FFCCC9}1.00 & \cellcolor[HTML]{FFCCC9}1.00 & \cellcolor[HTML]{FFCCC9}1.00 & \cellcolor[HTML]{FFCCC9}1.00 & \cellcolor[HTML]{FFCCC9}0.97 \\
& \textbf{15} & \cellcolor[HTML]{FFCCC9}1.00 & 0.70 & 0.83 & \cellcolor[HTML]{FFCCC9}1.00 & \cellcolor[HTML]{FFCCC9}1.00 & \cellcolor[HTML]{FFCCC9}1.00 & \cellcolor[HTML]{FFCCC9}1.00 & \cellcolor[HTML]{FFCCC9}1.00 & \cellcolor[HTML]{FFCCC9}1.00 & \cellcolor[HTML]{FFCCC9}1.00 & 0.95 \\
& \textbf{20} & \cellcolor[HTML]{FFCCC9}1.00 & \cellcolor[HTML]{FFCCC9}1.00 & 0.93 & \cellcolor[HTML]{FFCCC9}1.00 & 1.00 & \cellcolor[HTML]{FFCCC9}1.00 & \cellcolor[HTML]{FFCCC9}1.00 & \cellcolor[HTML]{FFCCC9}1.00 & \cellcolor[HTML]{FFCCC9}1.00 & \cellcolor[HTML]{FFCCC9}1.00 & \cellcolor[HTML]{FFCCC9}0.99 \\
& \textbf{25} & 0.97 & 0.93 & 0.83 & \cellcolor[HTML]{FFCCC9}1.00 & 0.97 & \cellcolor[HTML]{FFCCC9}1.00 & \cellcolor[HTML]{FFCCC9}1.00 & \cellcolor[HTML]{FFCCC9}1.00 & \cellcolor[HTML]{FFCCC9}1.00 & \cellcolor[HTML]{FFCCC9}1.00 & 0.97 \\
\multirow{-6}{*}{\textbf{\oneplusone}} & \textbf{29} & \cellcolor[HTML]{FFCCC9}1.00 & \cellcolor[HTML]{FFCCC9}1.00 & \cellcolor[HTML]{FFCCC9}1.00 & \cellcolor[HTML]{FFCCC9}1.00 & \cellcolor[HTML]{FFCCC9}1.00 & \cellcolor[HTML]{FFCCC9}1.00 & \cellcolor[HTML]{FFCCC9}1.00 & \cellcolor[HTML]{FFCCC9}1.00 & \cellcolor[HTML]{FFCCC9}1.00 & \cellcolor[HTML]{FFCCC9}1.00 & \cellcolor[HTML]{FFCCC9}1.00 \\
\bottomrule
\end{tabular}
}
\end{table}
Overall, \RS performs consistently poorly, exhibiting low AvgFDS values across all circuits and qubit counts, with overall averages between 0.20 and 0.30. These results indicate that unguided random exploration is largely ineffective for detecting faults within the given computational budget. This finding contrasts with the prior QOPS study, where random search was sufficient to detect all faults captured by existing testing approaches. The discrepancy arises from the nature of the faulty circuits considered in this work: unlike faulty circuits obtained through existing mutant generators~\cite{mutation,mutation2}, which introduce bigger changes in the circuit outputs, the faults introduced here are intentionally subtle and therefore require more guided exploration strategies.

\HC improves upon \RS, achieving moderate AvgFDS values with overall averages in the range of 0.40–0.48. While this demonstrates the benefit of local search, \HC still fails to consistently detect all faults. In contrast, \GA shows strong and stable performance across most qubit counts, with AvgFDS values typically exceeding 0.95. This improvement highlights the advantage of population-based search and genetic operators in effectively exploring the test-case space. Among all evaluated approaches, the \oneplusone achieves the best overall performance, attaining near-perfect or perfect AvgFDS values across almost all circuits and qubit counts, particularly for larger qubit counts.

At the individual circuit level, \GA and \oneplusone demonstrate consistently high detection rates across different circuits. In contrast, \RS exhibits substantial variability, frequently yielding zero or near-zero detection rates for several circuits. \HC shows mixed behavior, performing well on some circuits (e.g., Ae and Dj) while struggling on others (e.g., Graph and Real). Notably, the \oneplusone strategy displays the most robust behavior, achieving perfect detection (AvgFDS = 1.00) for nearly all circuits at higher qubit counts.

Table~\ref{tab:RQ1mds} results suggest a general trend of improved fault detection as the number of qubits increases. For \GA and \oneplusone, AvgFDS values approach 1.00 from 10 qubits onward, indicating good scalability with respect to circuit size. At first glance, this trend may appear counterintuitive, as increasing the number of qubits expands the search space exponentially and increases circuit complexity. However, we hypothesize that this behavior can be explained by the nature of quantum computation: as qubit count increases, the effects of injected faults tend to propagate more broadly through the circuit due to entanglement and superposition. As a result, even faults affecting a single qubit can influence a larger portion of the output state space, making them easier to detect but harder to localize as more qubits show faulty behavior. This phenomenon contrasts with classical programs, where faults and their effects are often localized to specific regions of the code.

\subsubsection{Similarity Analysis}
To verify whether fault effects spread across qubits as the qubit count increases, we conduct a similarity analysis of the generated failing test cases. Specifically, for each circuit and qubit count, we analyze the test cases produced across independent runs of the search strategy and compute their pairwise Jaccard similarity. This analysis allows us to assess whether different runs converge to similar failing test cases or whether multiple distinct test cases can expose the same fault. Formally, we compute the average pairwise similarity as

\begin{equation*}
\text{AvgSim} =
\frac{2}{N(N-1)}
\sum_{1 \le i < j \le N}
\text{Sim}(T_i, T_j),
\end{equation*}
where $N$ is the total number of runs and $T_i$, $T_j$ denote the failing test cases identified in the $i$th and $j$th runs, respectively, and $\text{Sim}(\cdot)$ is the Jaccard similarity defined as
\begin{equation*}
\text{Sim}(T_i,T_j) =
\begin{cases}
\dfrac{|T_i \cap T_j|}{|T_i \cup T_j|}, & \text{if } |T_i \cup T_j| > 0, \\[8pt]
-1, & \text{if } T_i = \varnothing \ \text{and} \ T_j = \varnothing .
\end{cases}
\end{equation*}

We report the average similarity across all run pairs. Since Jaccard similarity is undefined when both test cases are empty, we explicitly assign a similarity value of $-1$ in such cases. This design choice penalizes situations in which the search strategy fails to identify any failing test case, rather than incorrectly interpreting the absence of solutions as similarity. Consequently, negative average similarity values indicate inconsistency or ineffectiveness of the search process, rather than genuine dissimilarity among successful test cases (i.e, the search was not able to find a faulty test case, and we add -1 for comparison with the empty set).

\begin{figure}[!tb]
\centering
\includegraphics[width=0.6\columnwidth]{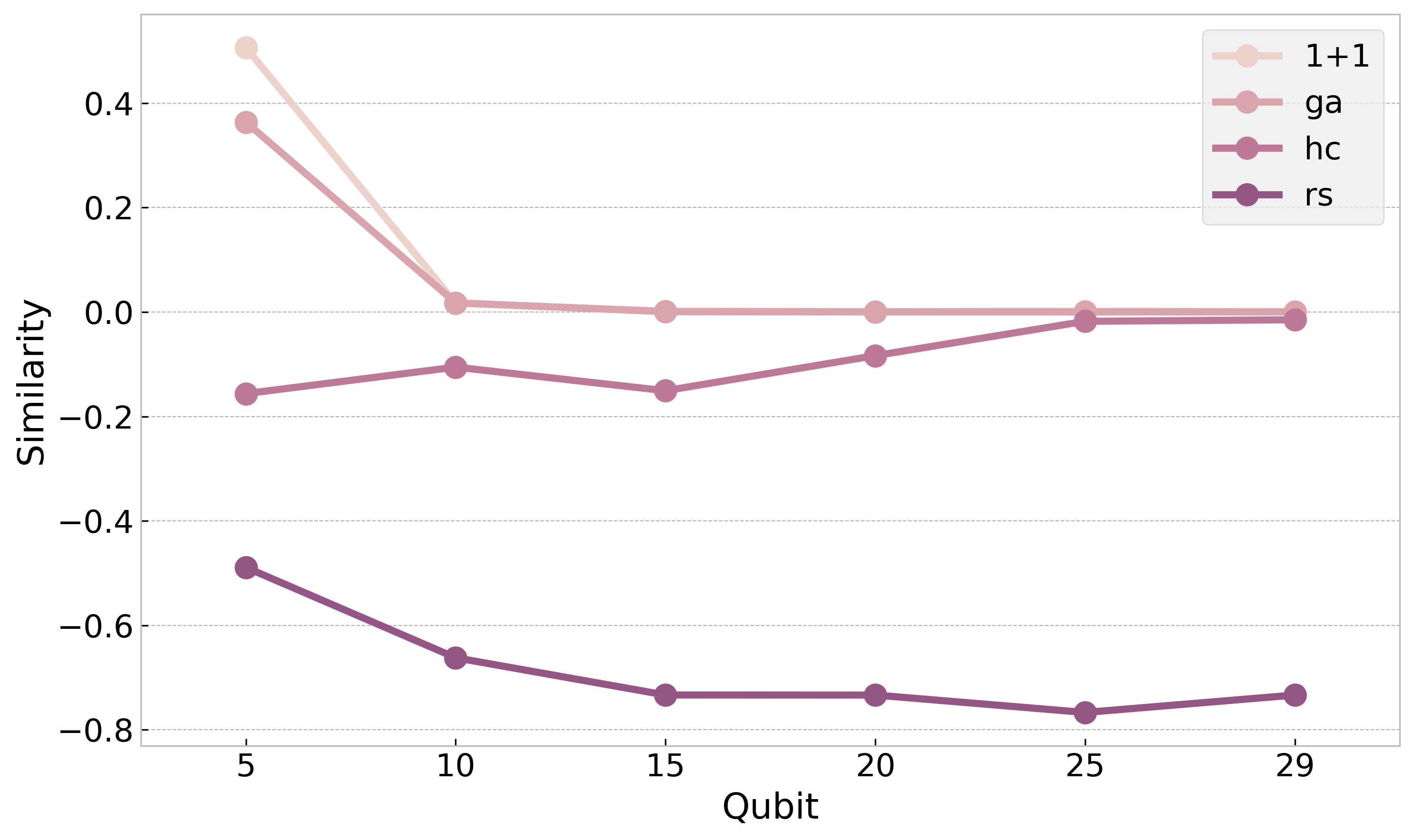}
\caption{RQ1 -- Average Jaccard similarity (AvgSim) of the test cases across ten runs for each search strategy and qubit count}
\label{fig:similarity}
\end{figure}

Figure~\ref{fig:similarity} presents the average test-case similarity across ten runs for each qubit count and search strategy.
The results show a clear trend: from 10 onward, both \GA and \oneplusone exhibit similarity values close to zero. This indicates that, although faults are consistently detected, the specific failing test cases differ substantially across runs. In other words, multiple distinct test cases are capable of exposing the same fault as the qubit count increases. This observation supports our earlier hypothesis that fault effects become more widely distributed in larger quantum circuits. While detecting faulty behavior becomes easier due to the proliferation of failing test cases, fault localization becomes more challenging. The relationship between faults and failing test cases transitions from near one-to-one in small circuits to one-to-many in larger circuits, complicating the identification of the specific circuit region responsible for the fault.

\subsubsection{Statistical Analysis}
To determine which search strategy is statistically best suited for \method, we conducted pairwise comparisons using the Wilcoxon signed-rank test and Cliff’s delta effect size across different runs, qubit counts, and circuits. The results of this analysis are summarized in Table~\ref{tab:RQ1stat}.
\begin{table}[!tb]
\centering
\caption{RQ1 -- Pairwise comparison of search strategies using the Wilcoxon signed-rank test and Cliff’s delta effect size. The strategy with superior performance is highlighted in bold. The Interpretation column reports the magnitude of the effect size, categorized as Large, Medium, Small, or Negligible according to guidelines~\cite{statistics2, recomended_effectsize}.}
\label{tab:RQ1stat}
\resizebox{0.7\columnwidth}{!}{%
\begin{tabular}{c|c|ccc}
\toprule
\textbf{Strategy A} & \textbf{Strategy B} & \textbf{P-value} & \textbf{Effect size} & \textbf{Interpretation} \\ \midrule
\RS & \textbf{\GA} & \textless{}0.05 & -0.78 & Large \\
\RS & \textbf{\HC} & \textless{}0.05 & -0.48 & Large \\
\RS & \textbf{\oneplusone} & \textless{}0.05 & -0.80 & Large \\
\textbf{\GA} & \HC & \textless{}0.05 & 0.79 & Large \\
\GA & \oneplusone & 0.394 & -0.08 & Negligible \\
\HC & \textbf{\oneplusone} & \textless{}0.05 & -0.80 & Large \\
\bottomrule
\end{tabular}
}
\end{table}
The statistical analysis indicates that \GA significantly outperforms both \RS and \HC ($p < 0.05$) with large effect sizes. In contrast, the comparison between \GA and \oneplusone does not reveal a statistically significant difference ($p = 0.394$) and yields a negligible effect size, suggesting that both strategies achieve comparable levels of effectiveness.

\begin{figure}[!tb]
\centering
% Row 1
\begin{subfigure}[t]{0.45\textwidth}
\centering
\includegraphics[width=\linewidth,height=4cm,keepaspectratio]{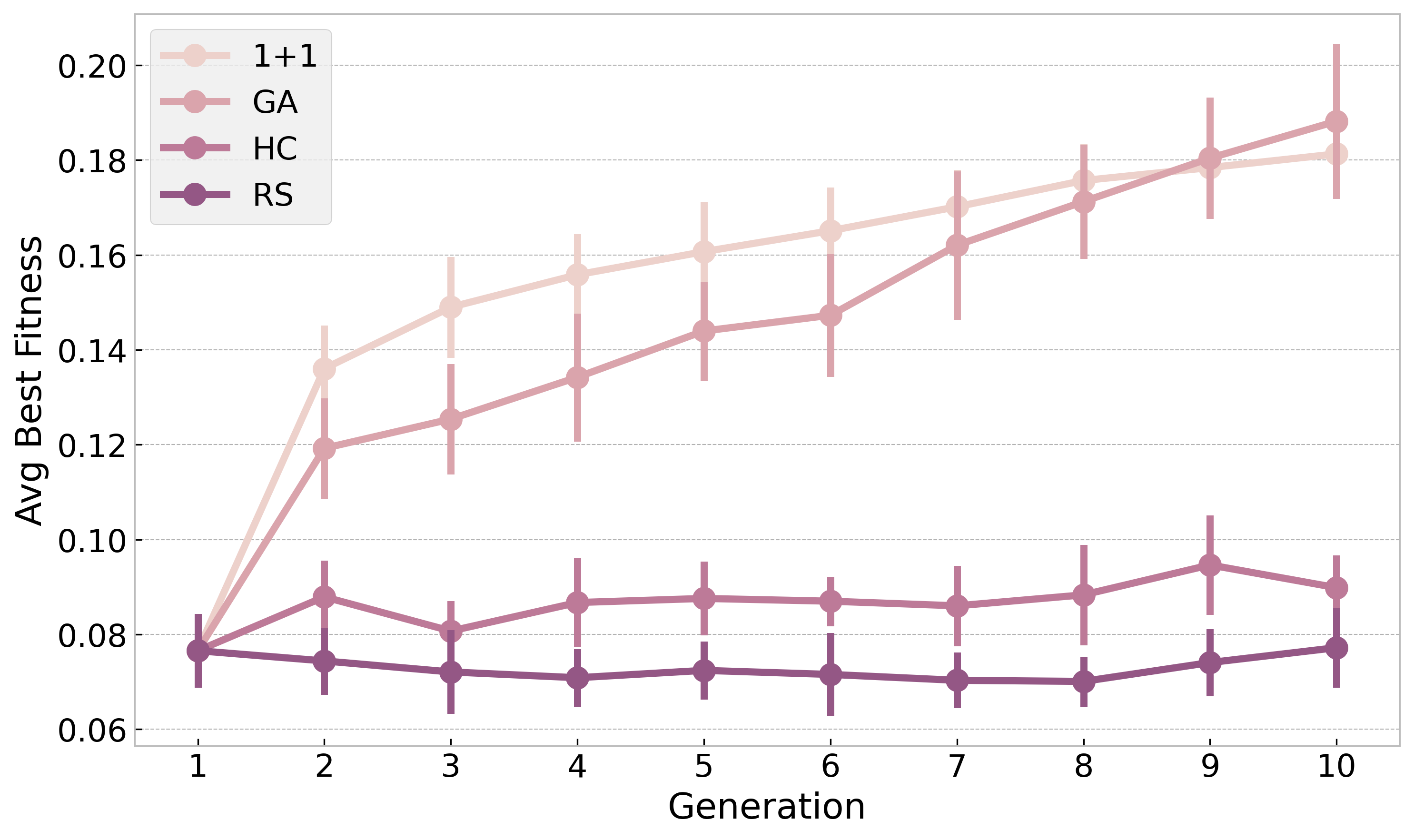}
\caption{5 Qubits}
\end{subfigure}
\hfill
\begin{subfigure}[t]{0.45\textwidth}
\centering
\includegraphics[width=\linewidth,height=4cm,keepaspectratio]{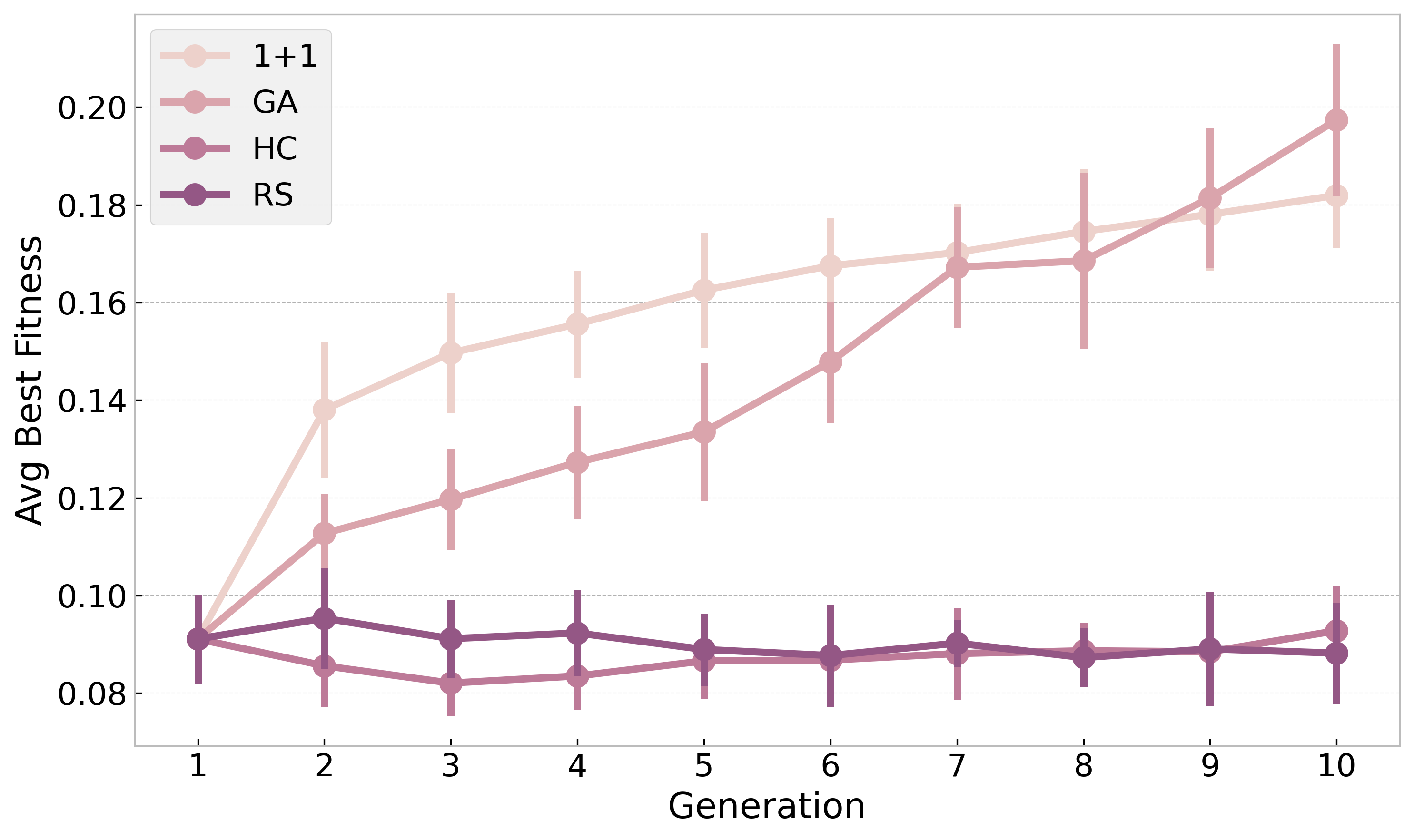}
\caption{10 Qubits}
\end{subfigure}

\vspace{0.4cm}

% Row 2
\begin{subfigure}[t]{0.45\textwidth}
\centering
\includegraphics[width=\linewidth,height=4cm,keepaspectratio]{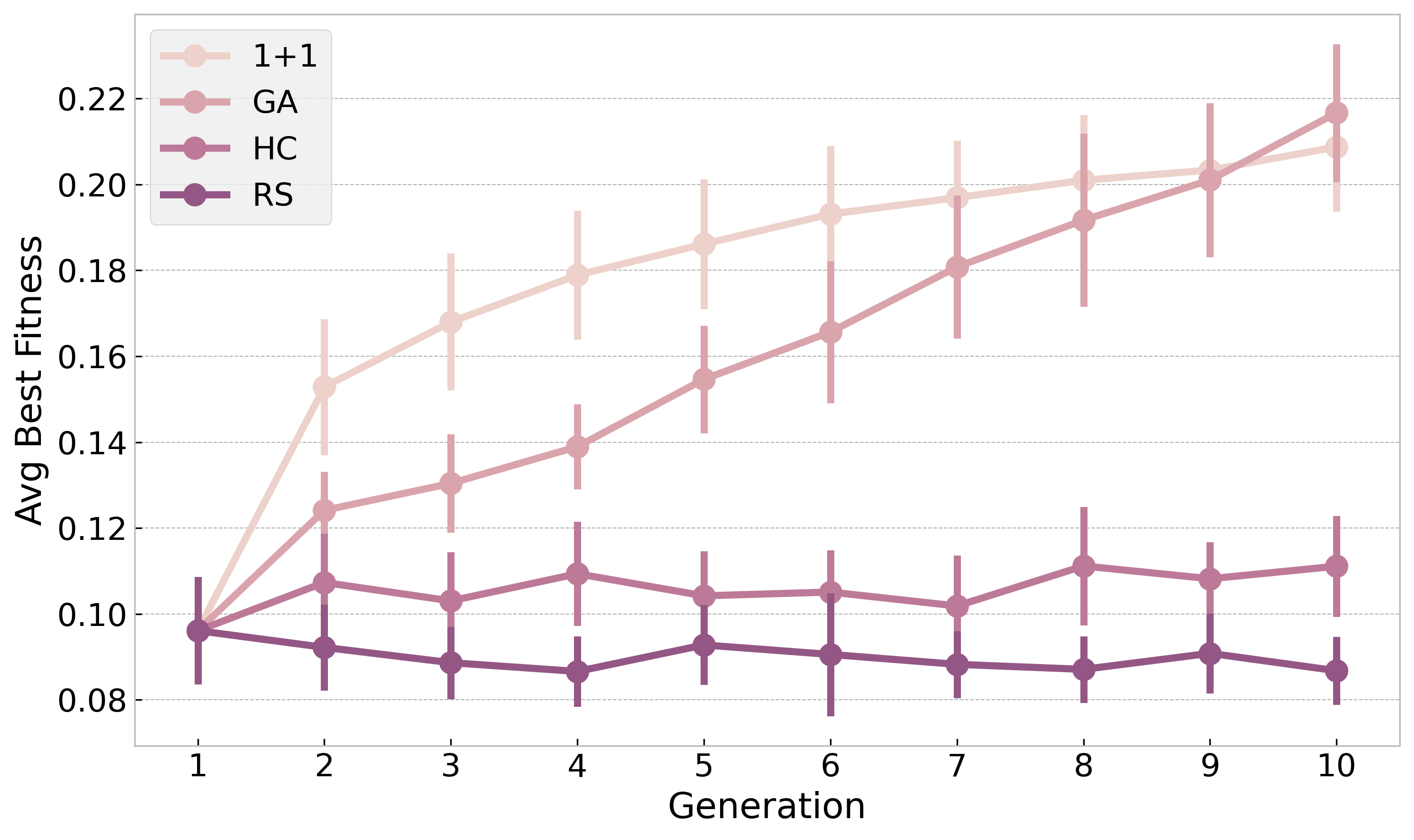}
\caption{15 Qubits}
\end{subfigure}
\hfill
\begin{subfigure}[t]{0.45\textwidth}
\centering
\includegraphics[width=\linewidth,height=4cm,keepaspectratio]{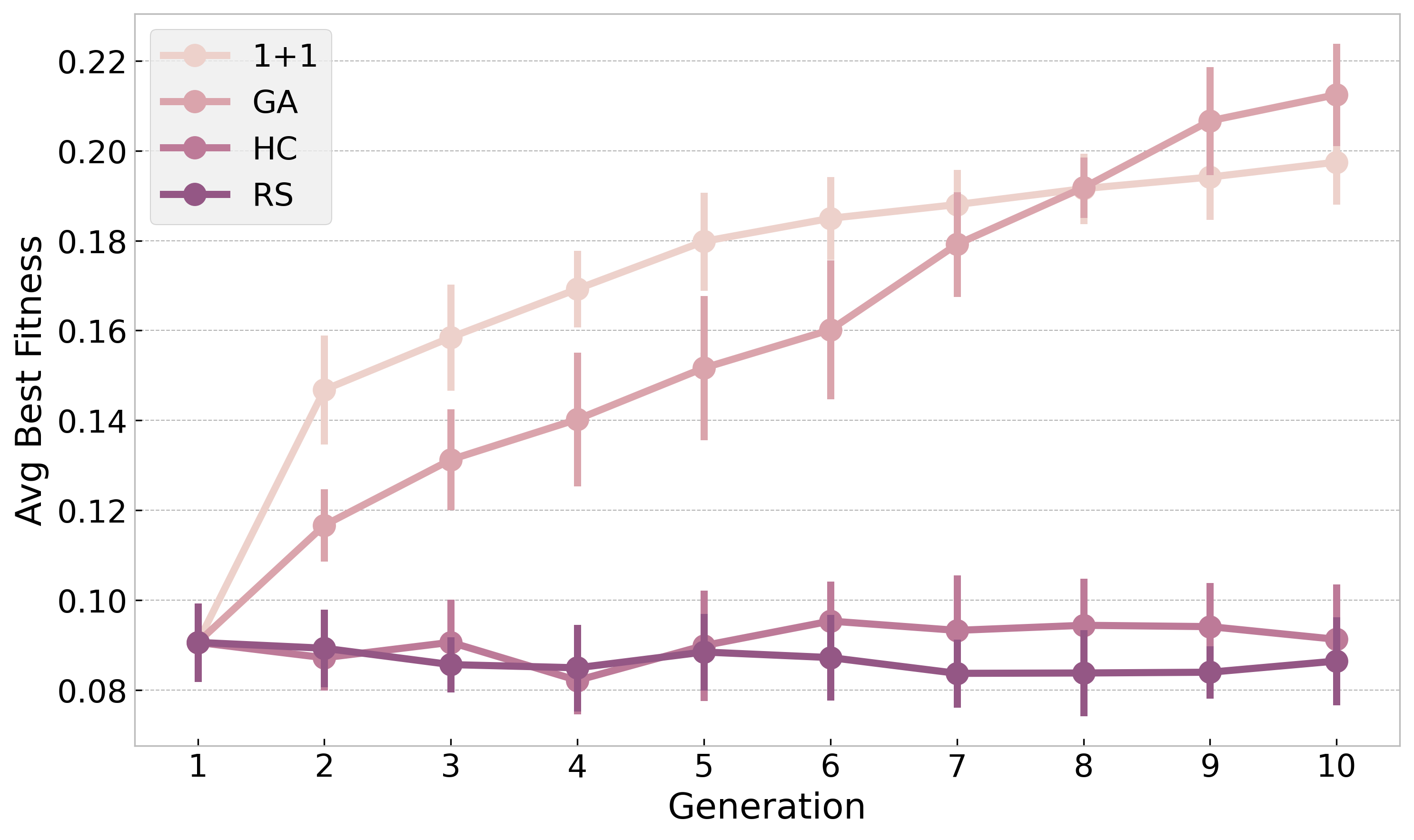}
\caption{20 Qubits}
\end{subfigure}

\vspace{0.4cm}

% Row 3
\begin{subfigure}[t]{0.45\textwidth}
\centering
\includegraphics[width=\linewidth,height=4cm,keepaspectratio]{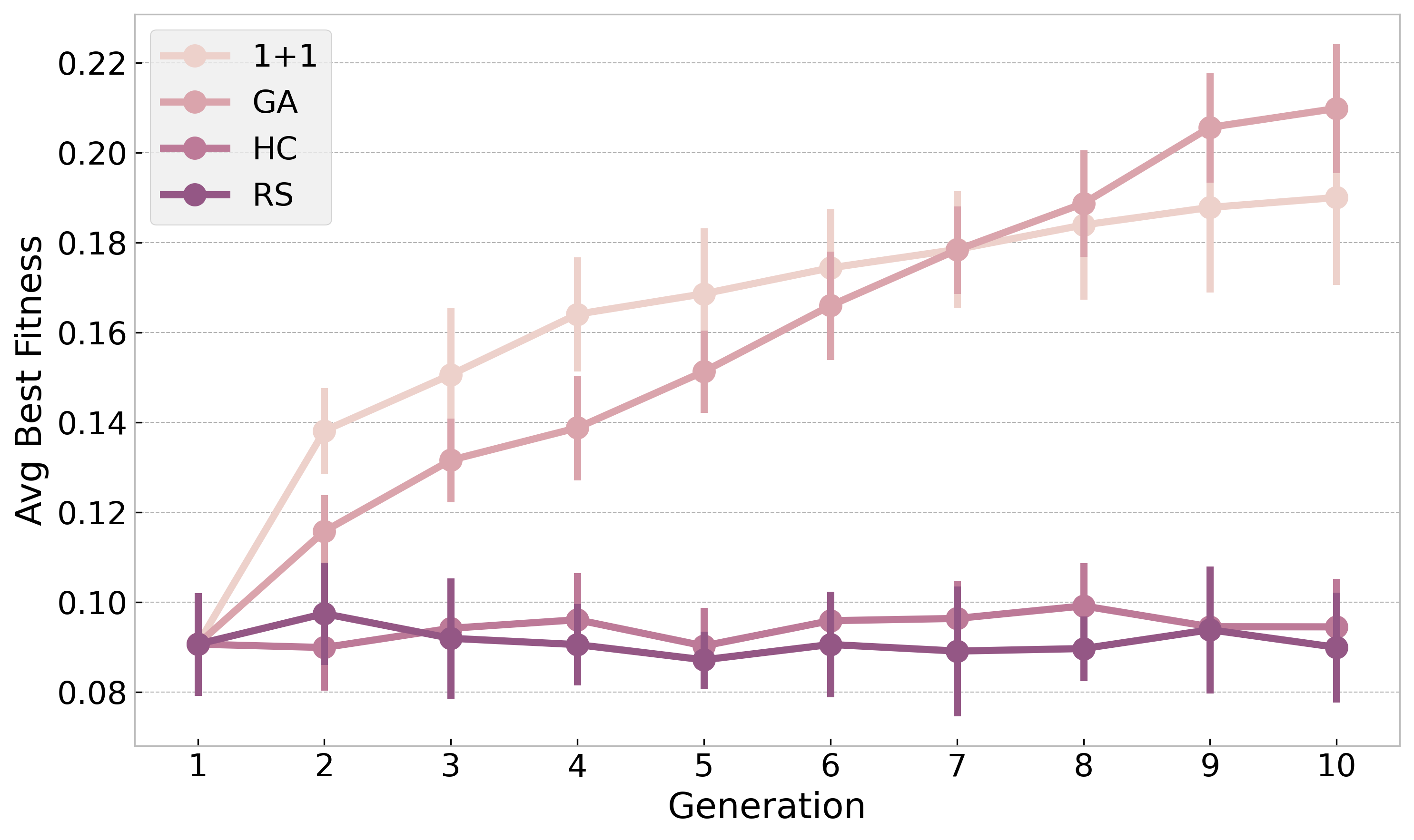}
\caption{25 Qubits}
\end{subfigure}
\hfill
\begin{subfigure}[t]{0.45\textwidth}
\centering
\includegraphics[width=\linewidth,height=4cm,keepaspectratio]{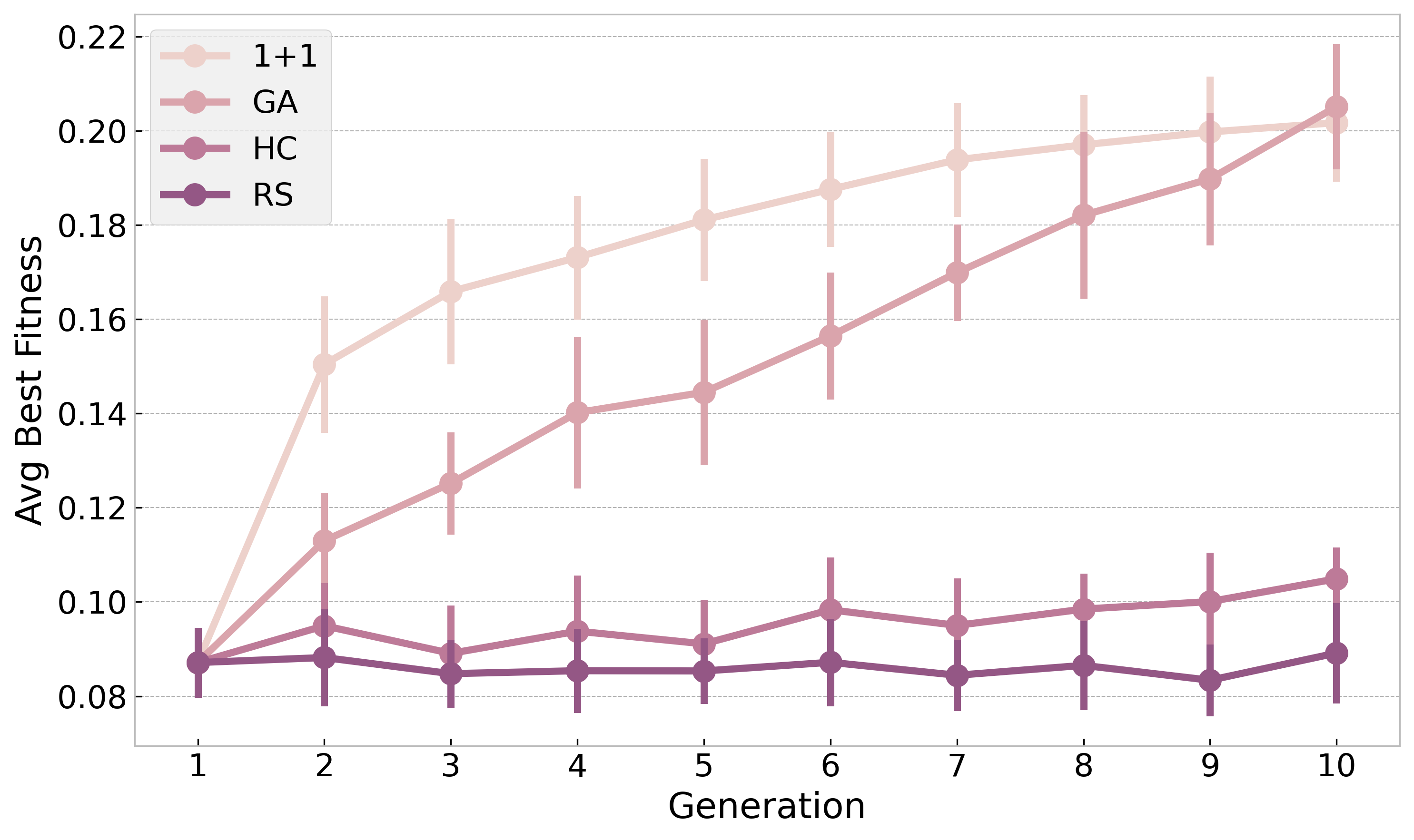}
\caption{29 Qubits}
\end{subfigure}
\caption{RQ1 -- Comparison of the average best fitness achieved by different search strategies across different qubit counts.}
\label{fig:fitness}
\end{figure}

To further investigate potential differences between the approaches (GA and \oneplusone beyond statistical significance, we analyze and visualize their fitness convergence behavior. Figure~\ref{fig:fitness} shows the average best fitness across 10 runs, aggregated over all circuits, as a function of generations for different qubit counts.
Because \GA is a population-based strategy, whereas \RS, \HC, and \oneplusone are iteration-based, a direct comparison would be inherently biased. To ensure fairness, we normalize computational effort by grouping 10 consecutive iterations of the iteration-based strategies into a single generation and picked maximum fitness value within the group, thereby aligning the computational cost of one generation across all approaches.

The results show that \GA consistently exhibits stable improvement in average best fitness across all qubit sizes and ultimately attains the highest fitness values. However, \oneplusone demonstrates faster convergence, often identifying solutions early in the search process. For all qubit counts, the best solutions of \GA and \oneplusone are very close to each other. This aligns with the statistical analysis that there is no significant difference in the final solution quality achieved by \GA and \oneplusone. However, in terms of only discovering a failing test case, \oneplusone is generally faster than \GA. Overall, these results suggest a trade-off between solution quality and convergence speed. \GA is more effective at optimizing the fitness value to its maximum, whereas \oneplusone is better suited when the primary goal is to rapidly identify a failing test case within a limited computational budget.

\begin{tcolorbox}[colback=blue!5!white,colframe=white,breakable]
\textbf{RQ1:} Compared to the previous QOPS approach, the \method is significantly more effective at identifying subtle faults. Among the evaluated strategies, \GA achieves higher-quality solutions in fitness optimization, while \oneplusone is faster at identifying failing test cases. Furthermore, \method demonstrates good scalability with increasing qubit counts, consistently finding failing test cases for bigger quantum circuits.
\end{tcolorbox}

\subsection{RQ2---Effect of Quantum Noise}
To address RQ2, we examine whether \method can reliably distinguish between faulty circuits and equivalent circuits in the presence of quantum noise. While a tolerance threshold of 1\% was sufficient to detect fault under ideal, noise-free simulation in RQ1, this threshold may no longer be appropriate under noisy execution. Consequently, we investigate how different threshold configurations affect the effectiveness of \method in noisy simulation. For this purpose, we evaluate \method using IBM’s noise models, both with and without error mitigation, to approximate the behavior of real quantum computers. Due to the substantial computational overhead of noisy simulation, especially when error mitigation is enabled, we restrict the RQ2 evaluation to circuits with up to 10 qubits. This restriction reflects the practical limitations of current noisy simulators~\cite{simulationlimit}. We further limit the search strategy to the \oneplusone strategy. As shown in RQ1, both \GA and \oneplusone achieved comparable fault-detection performance; however, \oneplusone converges faster, which allows us to run the evaluation multiple times for a more reliable analysis. As a first step, we execute \method on all benchmark circuits across the three IBM noise models without error mitigation and without applying a tolerance threshold. This configuration allows us to observe the impact of quantum noise on the fitness values produced by \method. We then repeat the same experimental setup with open-source error mitigation enabled.

\begin{figure}[!tb]
\includegraphics[width=1\columnwidth]{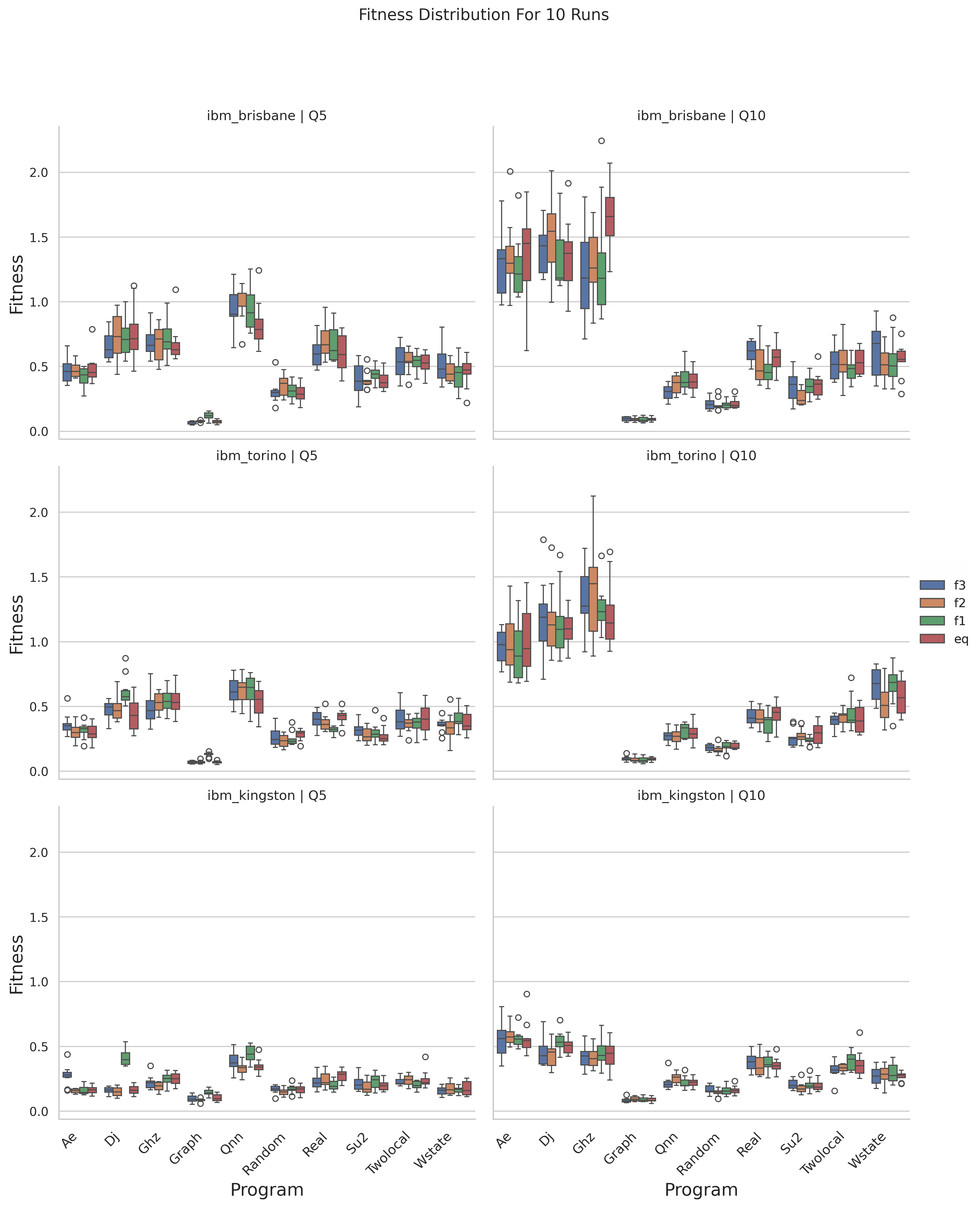}
\caption{RQ2 -- Fitness values obtained using the \oneplusone for each faulty circuit and qubit count under IBM noise simulation without error mitigation. Each box plot represents the distribution over 10 independent runs for a given circuit. f1, f2, and f3 are the faulty circuits, and eq represents the equivalent circuit.}
\label{fig:noisefitness}
\end{figure}

Figure~\ref{fig:noisefitness} shows the resulting fitness distributions of the best individuals of the 10 runs for all three noise models for 5- and 10-qubit circuits without error mitigation.
The results clearly demonstrate that fitness values are strongly influenced by both the circuit under test and the backend noise model. No uniform behavior emerges across benchmarks: each circuit exhibits a distinct fitness scale and distribution, and these distributions vary substantially across different backends.
%As a result, fitness values are not directly comparable across circuits, nor even across different backends for the same circuit.

More importantly, quantum noise significantly degrades fault detection. For every circuit and qubit count, the fitness distributions of equivalent and non-equivalent circuits overlap extensively. This overlap is consistent across all circuits and noise models, indicating that no single global threshold can reliably separate equivalent circuits from faulty ones. Although the Kingston noise exhibits slightly reduced variability compared to Brisbane and Torino, the absolute spread and overlap of the distributions remain large to support reliable threshold-based classification.

The impact of noise becomes even more pronounced for 10-qubit circuits, where the variability increases across most benchmarks. Consequently, any threshold tuned for a specific circuit and backend fails to generalize to other settings.

\begin{figure}[!tb]
\includegraphics[width=1\columnwidth]{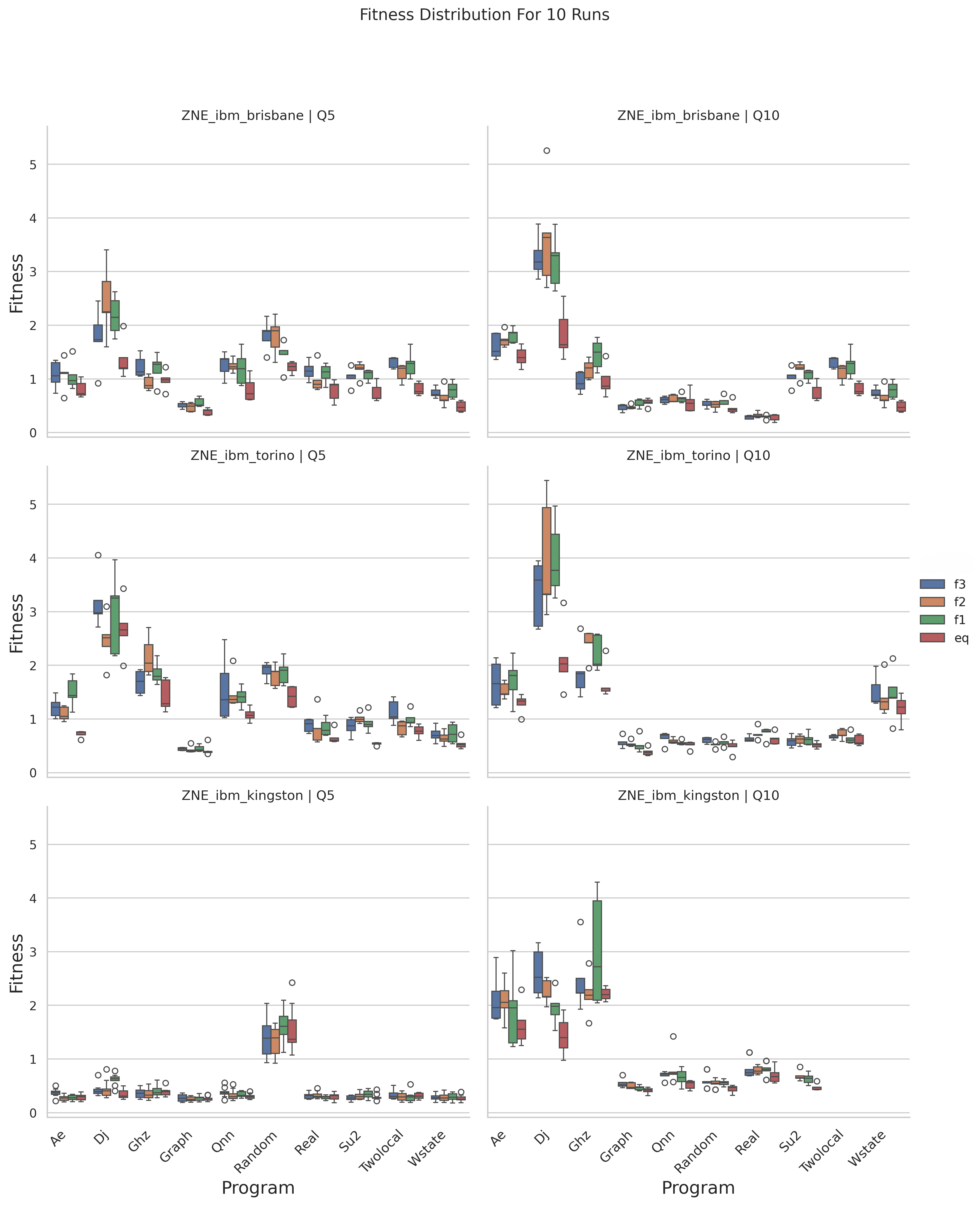}
\caption{RQ2 -- Fitness values obtained using the \oneplusone for each faulty circuit and qubit count under IBM noise simulation with ZNE error mitigation. Each box plot represents the distribution over 10 independent runs for a given circuit. f1, f2, and f3 are the faulty circuits, and eq represents the equivalent circuit.}
\label{fig:znefitness}
\end{figure}

Figure~\ref{fig:znefitness} presents the corresponding results with error mitigation enabled.
While error mitigation generally reduces the variance of the fitness distributions, the overall qualitative behavior remains unchanged. Fitness values continue to depend strongly on both the circuit and backend, and substantial overlap between equivalent and non-equivalent circuits persists across all configurations.

Overall, these findings indicate that, with open-source error mitigation, a single universal threshold is insufficient. Instead, a practical compromise is required: selecting threshold values that minimize false positives (i.e., correctly identifying equivalent circuits) for most circuits, while simultaneously reducing false negatives to maintain reasonably reliable fault detection under noisy execution.

To determine whether such a tolerance threshold exists, we construct a grid of 30 equally spaced threshold values, ranging from 0.1 to the maximum fitness value observed in RQ1 across all circuits (5.44). Following the classification framework defined in Section~\ref{metrics}, we compute precision, recall, and F1-score for each threshold configuration. Table~\ref{tab:RQ2noisecls} reports the resulting classification performance across the full threshold grid.
\begin{table}[!tb]
\caption{RQ2 -- Classification report for different threshold values across all circuits and noise models for 5- and 10-qubits.}
\label{tab:RQ2noisecls}
\centering
\setlength{\tabcolsep}{8pt}
%\resizebox{0.75\columnwidth}{!}{%
\begin{tabular}{c|ccc|cc}
\toprule
\textbf{Threshold} & \textbf{Precision} & \textbf{Recall} & \textbf{F1 Score} & \textbf{False Positives} & \textbf{False Negatives} \\ \midrule
0.1 & 0.75 & 1.0 & 0.64 & 58 & 0 \\
0.28 & 0.75 & 0.96 & 0.65 & 55 & 7 \\
0.47 & 0.78 & 0.8 & 0.68 & 39 & 34 \\
0.65 & 0.8 & 0.61 & 0.62 & 27 & 67 \\
0.84 & 0.82 & 0.49 & 0.56 & 19 & 89 \\
1.02 & 0.82 & 0.43 & 0.53 & 16 & 99 \\
1.21 & 0.81 & 0.35 & 0.47 & 14 & 113 \\
1.39 & 0.84 & 0.27 & 0.41 & 9 & 126 \\
1.57 & 0.86 & 0.21 & 0.36 & 6 & 136 \\
1.76 & 0.91 & 0.18 & 0.33 & 3 & 142 \\
1.94 & 0.89 & 0.14 & 0.29 & 3 & 149 \\
2.13 & 0.9 & 0.11 & 0.25 & 2 & 154 \\
2.31 & 1.0 & 0.08 & 0.22 & 0 & 159 \\
2.5 & 1.0 & 0.06 & 0.19 & 0 & 163 \\
2.68 & 1.0 & 0.05 & 0.17 & 0 & 165 \\
2.86 & 1.0 & 0.04 & 0.16 & 0 & 166 \\
3.05 & 1.0 & 0.04 & 0.16 & 0 & 166 \\
3.23 & 1.0 & 0.02 & 0.14 & 0 & 169 \\
3.42 & 1.0 & 0.01 & 0.11 & 0 & 172 \\
3.6 & 1.0 & 0.01 & 0.11 & 0 & 172 \\
3.79 & 1.0 & 0.01 & 0.11 & 0 & 172 \\
3.97 & 1.0 & 0.01 & 0.11 & 0 & 172 \\
4.15 & - & - & - & 0 & 173 \\
4.34 & - & - & - & 0 & 173 \\
4.52 & - & - & - & 0 & 173 \\
4.71 & - & - & - & 0 & 173 \\
4.89 & - & - & - & 0 & 173 \\
5.08 & - & - & - & 0 & 173 \\
5.26 & - & - & - & 0 & 173 \\
5.44 & - & - & - & 0 & 173 \\
\bottomrule
\end{tabular}
%}
\end{table}

The results reveal a clear trade-off between fault detection capability and equivalence identification as the threshold varies. At very low threshold values (i.e., 0.10–0.28), recall is near-perfect (1.00 and 0.96, respectively), indicating that almost all faulty circuits are correctly detected. However, this high recall comes at the cost of a large number of false positives (58 and 55), meaning that many equivalent circuits are incorrectly classified as faulty.

As the threshold increases, the number of false positives steadily decreases. This trend is particularly evident in the range between 0.47 and 0.84, where false positives drop from 39 to 19 and precision increases from 0.78 to 0.82. However, this improvement is accompanied by a substantial reduction in recall (from 0.80 to 0.49) and a corresponding increase in false negatives. Although the F1-score reaches its maximum value of 0.68 at a threshold of 0.47, indicating the best balance between precision and recall, this performance remains unsatisfactory in practice due to the still large number of both false positives and false negatives.

For higher threshold values, most faulty circuits are classified as equivalent, resulting in minimal false positives but severely compromised fault-detection capability. For thresholds greater than 4.15, no positive predictions are produced at all, rendering precision, recall, and F1-score undefined while all faulty circuits are misclassified. Overall, these findings confirm that, with open-source error mitigation and IBM noise models, no single universal threshold can simultaneously optimize precision and recall across all circuits.

The results of RQ2 contrast with those reported in the previous QOPS study~\cite{qops} conducted under noisy conditions. In that work, circuit-specific thresholds were derived using standard error estimates combined with error mitigation, enabling effective discrimination between equivalent and non-equivalent circuits under IBM noise models. The discrepancy between the two studies arises primarily from the nature of the circuits. In the previous work, mutation operators from mutation tools~(\cite{mutation, mutation2}) produced faulty circuits whose outputs differed substantially from the original circuits. Such faults remain distinguishable even under noisy execution using circuit-specific thresholds. In contrast, the faults considered in this study are manually constructed, subtle faults, introducing careful changes that lead to minimal deviations in the output distributions.

These subtle output differences are more easily masked by quantum noise, making discrimination significantly more challenging. As a result, the available open-source noise-handling approach is ineffective in this setting. This highlights the need for improved error-mitigation methods to detect fine-grained faults under noisy conditions.

\begin{tcolorbox}[colback=blue!5!white,colframe=white,breakable]
\textbf{RQ2:} Under IBM noise models with open-source error mitigation, \method exhibits an inherent trade-off when detecting subtle faults. Low tolerance thresholds yield high recall but incur a large number of false positives, whereas higher thresholds reduce false positives at the expense of missing most faulty circuits. As a result, no universal threshold can reliably distinguish faulty from equivalent circuits, underscoring the need for more effective open-source error-mitigation techniques.
\end{tcolorbox}

\subsection{RQ3---Applicability on Real Quantum Computers}
The primary objective of RQ3 is to evaluate the applicability of \method across diverse quantum computing architectures, thereby assessing whether the measurement-based test cases defined by \method are compatible with current real quantum computers. To this end, we evaluate \method on IBM’s real quantum computers, IQM’s real quantum computer, and Quantinuum’s quantum emulator (as the promotional access is insufficient to execute on the real computer), covering multiple hardware technologies beyond a single vendor ecosystem.

For the RQ3 evaluation, we utilized IBM’s 156-qubit quantum computer through IBM’s Quantum Credit Program, which provides selected research groups with limited access (typically 5–12 QPU hours) to real quantum hardware. In addition, we employed IQM’s 56-qubit quantum computer and Quantinuum’s QNexus platform through similar collaborative access arrangements. Despite having promotional access to these systems, the high cost of real quantum execution makes it infeasible to evaluate the full \method configuration across all qubit sizes and multiple independent runs, as done in RQ1.

Consequently, we restrict the RQ3 evaluation to configurations that maximize insights while efficiently utilizing available QPU resources. Specifically, for both IBM and IQM quantum computers, we evaluate \method using the \oneplusone evolutionary algorithm on 29-qubit circuits, which represents the largest qubit count considered in RQ1 and allows execution across all benchmark circuits and their corresponding faulty circuits within the available QPU budget. Even with this restriction, performing complete evolutionary search runs on real quantum computers remains infeasible. Therefore, instead of rerunning the full search process, we reuse the failing test cases identified in RQ1 for the 29-qubit count. These test cases, obtained across 10 independent runs for each faulty circuit, are executed directly on real quantum hardware. To enable a similar classification analysis analogous to RQ2, each test case is also executed on the corresponding equivalent (non-faulty) circuit.

IBM quantum computers provide built-in ZNE error mitigation and additionally support probabilistic error amplification (PEA)~\cite{pea} to enhance the effectiveness of ZNE. We enable these built-in error mitigation techniques for IBM hardware to achieve maximum noise reduction. Notably, we did not use PEA-enhanced ZNE in RQ2, as we were unable to find an open-source implementation of this method.

For the IQM quantum computer, we follow the same experimental configuration as IBM in terms of qubit count and search strategy. However, IQM does not currently provide built-in error mitigation comparable to IBM’s ZNE, and porting or implementing custom error mitigation techniques for real hardware is beyond the scope of this study. Therefore, all IQM results are obtained without error mitigation.

Table~\ref{tab:RQ3noisecls} reports the classification results for the 29-qubit circuits and their corresponding equivalent and non-equivalent circuits executed on IBM and IQM real quantum computers.
\begin{table}[!tb]
\caption{RQ3 -- Classification report for different threshold values across all circuits on IBM and IQM for 29-qubit failing test cases. The best threshold from our point of view is highlighted in red.}
\label{tab:RQ3noisecls}
\setlength{\tabcolsep}{3pt}
\centering
%\resizebox{0.8\textwidth}{!}{%
\begin{tabular}{c|ccccc|cccll}
\toprule
\multirow{2}{*}{\textbf{Threshold}} & \multicolumn{5}{c|}{\textbf{IBM}} & \multicolumn{5}{c}{\textbf{IQM}} \\
\cmidrule{2-6} \cmidrule{7-11}
& \textbf{Precision} & \textbf{Recall} & \textbf{F1 Score} & \textbf{FP} & \textbf{FN} & \textbf{Precision} & \textbf{Recall} & \textbf{F1 Score} & \multicolumn{1}{c}{\textbf{FP}} & \multicolumn{1}{c}{\textbf{FN}} \\
\midrule
0.1 & 0.88 & 0.77 & 0.83 & 3 & 7 & 0.52 & 0.78 & 0.51 & 19 & 9 \\
\cellcolor[HTML]{FFCCC9}\textbf{0.28} & \cellcolor[HTML]{FFCCC9}\textbf{1} & \cellcolor[HTML]{FFCCC9}\textbf{0.47} & \cellcolor[HTML]{FFCCC9}\textbf{0.71} & \cellcolor[HTML]{FFCCC9}\textbf{0} & \cellcolor[HTML]{FFCCC9}\textbf{16} & 0.52 & 0.63 & 0.51 & 16 & 13 \\
0.47 & 1 & 0.33 & 0.62 & 0 & 20 & 0.48 & 0.56 & 0.48 & 16 & 15 \\
0.65 & 1 & 0.27 & 0.58 & 0 & 22 & 0.46 & 0.44 & 0.46 & 14 & 18 \\
0.84 & 1 & 0.27 & 0.58 & 0 & 22 & 0.5 & 0.41 & 0.5 & 11 & 19 \\
1.02 & 1 & 0.17 & 0.5 & 0 & 25 & 0.5 & 0.33 & 0.49 & 9 & 21 \\
1.2 & 1 & 0.17 & 0.5 & 0 & 25 & 0.5 & 0.3 & 0.48 & 8 & 22 \\
1.39 & 1 & 0.1 & 0.44 & 0 & 27 & 0.5 & 0.3 & 0.48 & 8 & 22 \\
1.57 & 1 & 0.1 & 0.44 & 0 & 27 & 0.5 & 0.26 & 0.47 & 7 & 23 \\
1.76 & 1 & 0.1 & 0.44 & 0 & 27 & 0.54 & 0.26 & 0.48 & 6 & 23 \\
1.94 & 1 & 0.1 & 0.44 & 0 & 27 & 0.5 & 0.22 & 0.46 & 6 & 24 \\
2.13 & 1 & 0.1 & 0.44 & 0 & 27 & 0.5 & 0.22 & 0.46 & 6 & 24 \\
2.31 & 1 & 0.07 & 0.4 & 0 & 28 & 0.5 & 0.22 & 0.46 & 6 & 24 \\
2.49 & 1 & 0.07 & 0.4 & 0 & 28 & 0.5 & 0.22 & 0.46 & 6 & 24 \\
2.68 & 1 & 0.07 & 0.4 & 0 & 28 & 0.5 & 0.22 & 0.46 & 6 & 24 \\
2.86 & 1 & 0.07 & 0.4 & 0 & 28 & 0.5 & 0.19 & 0.44 & 5 & 25 \\
3.05 & 1 & 0.07 & 0.4 & 0 & 28 & 0.5 & 0.19 & 0.44 & 5 & 25 \\
3.23 & 1 & 0.03 & 0.37 & 0 & 29 & 0.5 & 0.19 & 0.44 & 5 & 25 \\
3.41 & 1 & 0.03 & 0.37 & 0 & 29 & 0.5 & 0.15 & 0.43 & 4 & 26 \\
3.6 & 1 & 0.03 & 0.37 & 0 & 29 & 0.43 & 0.11 & 0.4 & 4 & 27 \\
3.78 & 1 & 0.03 & 0.37 & 0 & 29 & 0.5 & 0.11 & 0.41 & 3 & 27 \\
3.97 & 1 & 0.03 & 0.37 & 0 & 29 & 0.5 & 0.11 & 0.41 & 3 & 27 \\
4.15 & 1 & 0.03 & 0.37 & 0 & 29 & 0.5 & 0.07 & 0.39 & 2 & 28 \\
4.34 & 1 & 0.03 & 0.37 & 0 & 29 & 0.5 & 0.07 & 0.39 & 2 & 28 \\
4.52 & 1 & 0.03 & 0.37 & 0 & 29 & 1 & 0.04 & 0.37 & 0 & 29 \\
4.7 & - & - & - & 0 & 30 & - & - & - & 0 & 30 \\
4.89 & - & - & - & 0 & 30 & - & - & - & 0 & 30 \\
5.07 & - & - & - & 0 & 30 & - & - & - & 0 & 30 \\
5.26 & - & - & - & 0 & 30 & - & - & - & 0 & 30 \\
5.44 & - & - & - & 0 & 30 & - & - & - & 0 & 30 \\
\bottomrule
\end{tabular}
%}
\end{table}
For IQM, the results closely mirror the behavior observed in RQ2 under noisy simulation. Subtle output differences are easily masked by quantum noise, and no threshold is able to simultaneously optimize both precision and recall. As in RQ2, aggressive thresholds lead to high recall but many false positives, while conservative thresholds suppress false positives at the expense of missing most faulty circuits. This confirms that, in the absence of effective error mitigation, \method struggles to reliably distinguish faulty from equivalent circuits on real hardware.

In contrast, the IBM results show a markedly different trend. Threshold values of 0.1 and 0.28 perform reasonably well, with 0.1 corresponding to the same threshold used in the ideal simulations of RQ1. At a threshold of 0.1, the classification yields three false positives and seven false negatives. This setting is suitable when the primary objective is to detect faulty circuits, even at the cost of occasionally misclassifying equivalent ones. However, from a software testing perspective, false positives are particularly undesirable, as misclassifying equivalent circuits as faulty undermines the reliability of the testing process.

Accordingly, a threshold of 0.28 emerges as the most suitable compromise: it produces zero false positives while still identifying a subset of faulty circuits, albeit with some false negatives. In our view, tolerating false negatives is preferable to introducing false positives, making 0.28 the most appropriate threshold for applying \method on IBM’s real quantum computer in this setting.

Notably, the IBM results in RQ3 stand in contrast to those of RQ2. In RQ2, no global threshold could be identified under noisy simulation, even with open-source error mitigation enabled. In RQ3, despite executing on real quantum computer (which might be noisier than simulation) a usable threshold becomes feasible. The key differentiating factor is error mitigation. While IQM results remain similar to RQ2 due to the absence of error mitigation, IBM’s hardware benefits from a significantly more advanced mitigation pipeline. IBM’s implementation of ZNE combined with PEA is substantially more effective than the open-source ZNE approach used in RQ2, enabling meaningful noise reduction even for 29-qubit circuits. This level of mitigation is sufficient for \method to remain applicable on real quantum computers, underscoring the critical importance of strong, hardware-integrated error mitigation.

For Quantinuum, the available promotional access budget did not permit execution of even the restricted configuration used for IBM and IQM. However, Quantinuum’s 20-qubit emulator allowed us to execute the full (\oneplusone) \method for 10 runs within the allocated budget. These experiments were conducted to verify whether the test cases generated by \method are executable on Quantinuum’s quantum computer architecture. All executions completed successfully, confirming architectural compatibility. Nevertheless, the results again resembled those of RQ2: without error mitigation enabled, no reliable threshold could be identified.

Overall, these findings demonstrate that \method is architecturally portable and executable across diverse quantum computing platforms. However, effective error mitigation is essential for reliable test assessment on real quantum computers.

\begin{tcolorbox}[colback=blue!5!white,colframe=white,breakable]
\textbf{RQ3:} \method is architecturally portable and can be executed across different quantum computing platforms, including IBM, IQM, and Quantinuum. However, its practical effectiveness on real quantum computers essentially depends on the availability of strong error mitigation mechanisms to enable reliable test assessment.
\end{tcolorbox}

\section{Threats to validity}\label{sec:threats}
This section discusses the threats to validity across four established categories~\cite{threats}: construct validity, internal validity, conclusion validity, and external validity.

\subsection{Construct Validity}
Construct validity refers to how well a measurement captures the theoretical concept it is intended to evaluate. One threat to construct validity in our study arises from the choice of metrics used to assess the effectiveness of \method. Since the primary goal of \method is to detect faults in quantum circuits, we evaluate its effectiveness using the average fault detection score. Under quantum noise, however, a testing approach must also reliably distinguish between faulty and non-faulty (equivalent) circuits. Accordingly, the comparative effectiveness of testing approaches depends on the number of false positives and false negatives produced during assessment. To capture this behavior, we employ precision, recall, and F1-score as evaluation metrics. Low precision and recall indicate higher rates of false positives and false negatives, respectively, while the F1-score provides a balanced measure of overall classification performance.

Another potential threat to construct validity concerns the choice of error mitigation techniques and the quantum hardware platforms used in the evaluation. Different quantum computers exhibit varying noise characteristics, and error mitigation methods may have different effectiveness depending on the circuit structure and backend. To address this, we evaluate \method across multiple quantum computing platforms, including IBM, IQM, and Quantinuum, to demonstrate its broad applicability.

With respect to error mitigation, each method incurs different computational overheads and exhibits varying effectiveness. We selected Zero-Noise Extrapolation (ZNE) due to its wide adoption, open-source availability, and use in industrial settings such as IBM’s quantum systems. Given the limited access to real quantum hardware, ZNE was the only practically feasible error mitigation approach for our study. We employ two variants of ZNE: an open-source implementation provided by Mitiq~\cite{mitiq}, and IBM’s ZNE with PEA enhancement implementation used on their real quantum computers.

\subsection{Internal Validity}
Internal validity concerns the extent to which an experiment can establish a causal relationship between the independent and dependent variables. One potential threat to internal validity in our study arises from the tolerance parameter of \method. This parameter determines whether the fitness value of a test case is sufficient to classify it as failing. The appropriate value of this tolerance depends on the circuit under test and may require domain knowledge to be set correctly. In our experiments under ideal (noise-free) conditions, we selected the tolerance value to match the threshold used in the previous QOPS approach, namely 1\%, to ensure a fair and direct comparison between QOPS and \method.

For evaluations involving noisy simulations and real quantum computers, the choice of tolerance becomes more challenging. Quantum noise affects different circuits differently, and error mitigation techniques introduce their own circuit-dependent error margins. To mitigate this threat, we performed a grid search over a range of tolerance values and evaluated \method under multiple tolerance settings.

\subsection{Conclusion Validity}
Conclusion validity concerns the extent to which the statistical conclusions drawn from an experiment are sound and reliable. One potential threat to conclusion validity in our study arises from the choice of hyperparameters for both the search strategies and the ZNE error mitigation method.

For the search strategies (\GA, \HC, and \oneplusone), we employed the built-in hyperparameter tuning functionality provided by the Mealpy library to identify suitable configurations. Specifically, we conducted a preliminary experiment on 5-qubit circuits, in which \method was executed with different search strategies while systematically tuning their hyperparameters. This process resulted in the selection of 10 generations with a population size of 40 for \GA, and 140 iterations for both \HC and \oneplusone. The tuning process also identified appropriate values for other parameters, such as selection strategy and mutation rate.

Regarding the hyperparameters of ZNE, multiple configurations are available in the Mitiq library~\cite{mitiq}. We used Mitiq’s built-in configuration optimizer to determine an effective open-source ZNE setup for our benchmark circuits. This optimization selected noise-scaling factors of 1.0, 2.0, and 3.0, combined with polynomial extrapolation of order three as the factory method.

For executions on IBM’s real quantum computers, we selected the best option available, namely ZNE enhanced with Probabilistic Error Amplification (PEA). This choice minimizes the risk that suboptimal error mitigation configurations could bias the experimental results and strengthens the statistical validity of our conclusions.

\subsection{External Validity}
External validity refers to the extent to which the findings of this study can be generalized to other datasets, circuits, and quantum computing platforms. One potential threat to external validity lies in the selection of quantum circuits used for test assessment, as a limited or biased set of circuits could restrict the generalizability of the results. To mitigate this threat, we selected benchmark circuits from the widely adopted MQT benchmark suite~\cite{mqt}, which includes representative instances of commonly used quantum circuits across different application domains.

Another potential threat concerns the number of qubits evaluated. As the number of qubits increases, the complexity of quantum circuits grows substantially, leading to an exponential expansion of the state space. Compared to our previous study, we extended the evaluation to circuits up to 29 qubits, which represents the practical upper limit for current quantum simulators within reasonable execution times~\cite{speed}. Also, we evaluated \method on IBM, IQM, and Quantinuum quantum computers, demonstrating its applicability across different architectures.
%To the best of our knowledge, this study is among the few that evaluate test assessment techniques at such a large scale (up to 29 qubits) and additionally demonstrate the approach's execution on diverse real quantum computers at this scale.

\section{Related Work}\label{sec:related}
\subsection{Search-Based Software Testing}
Search-based software testing (SBST) is a well-established research area that applies metaheuristic optimization techniques to automate and enhance software testing processes. A comprehensive survey by~\cite{ssbsetesting1} presents a systematic account of SBST’s development, discussing its foundational concepts, historical evolution, and primary application domains, including test data generation, structural testing, and non-functional testing, as well as highlighting open research challenges. From a complementary standpoint, the study in~\cite{harman2009theoretical, HarmanICST2015} provides an empirical synthesis of the field, examining how different search strategies (evolutionary and non-evolutionary), along with fitness functions and problem formulations, perform across different testing scenarios. In quantum software testing, search-based approaches have also been explored: QuSBT~\cite{search} employs search strategies to generate effective test cases, while MuTG~\cite{mutation} adopts a multi-objective optimization strategy to minimize the number of test cases while maximizing mutant detection. In this study, we employ a search strategy to systematically generate test cases, similar in spirit to QuSBT and MuTG. The key difference lies in our definition of test cases and the new expectation value–based test oracle. Existing search-based quantum testing approaches rely on simple computational-basis inputs and require a full program specification and statistical test oracle. In contrast, \method defines test cases at the measurement level and employs an expectation value-based oracle, making a direct comparison with existing search-based testing techniques inapplicable.

\subsection{Quantum Software Testing}
Several studies have directed their research efforts toward quantum software testing, resulting in a variety of techniques for validating quantum circuits, assessing test suite adequacy, and defining appropriate test oracles. Different tools have been proposed to support these tasks. For example, Quito~\cite{quitoASE21tool} focuses on testing quantum circuits using input–output coverage criteria. Mutation testing support is provided by tools such as Muskit~\cite{mutation} and QMutpy~\cite{mutation2}. Other techniques include QuCAT~\cite{Combinatorial}, which applies combinatorial testing principles, and QuanFuzz~\cite{fuzz}, which adapts fuzzing techniques for quantum circuit testing. To evaluate the quality of test suites, several adequacy criteria have been introduced in the literature. Commonly used metrics include input–output coverage~\cite{coverage} and equivalence class partitioning~\cite{LongTOSEM2024}. With respect to test oracles, existing approaches typically fall into two categories. The first relies on statistical hypothesis testing to compare the output distributions of quantum circuits~\cite{coverage, Huang2019}. The second category comprises property-based testing methods~\cite{property,property2}, where violations of predefined circuit properties are treated as indicators of faults. In addition, several bug detection frameworks and benchmarks have been developed~\cite{bug1,bug2,MiranskyyICSE2020}, including Bugs4Q~\cite{bug3}, which provides a curated set of real-world bugs extracted from Qiskit programs.

A major challenge for quantum software testing stems from the inherently noisy nature of current quantum hardware~\cite{noise}. Noise arises from multiple sources, such as environmental interference (e.g., magnetic fields and radiation), qubit crosstalk, and gate calibration errors~\cite{noisesource}, and it significantly undermines the reliability of circuit outputs. As a result, distinguishing genuine software faults from noise-induced errors on real quantum computers becomes particularly difficult. To address this problem, a range of quantum error mitigation techniques has been proposed~\cite{qem}, most of which aim to reduce the impact of noise on expectation values.

Despite these advances, only a limited number of testing approaches currently support execution on real quantum hardware. QOIN~\cite{qoin} trains a machine-learning (ML) model using a small set of statistical features to mitigate noise effects. Qraft~\cite{qraft} extends this idea by introducing features that generalize across multiple circuits, while Q-LEAR~\cite{qlear} proposes a depth-cut-error feature to overcome Qraft’s tendency to overestimate noise in circuit outputs.

Nevertheless, existing quantum testing methods exhibit important limitations. Most approaches are incompatible with search-based and optimization-oriented quantum circuits and require complete program specifications, which are often unavailable in practice. Apart from QOIN~\cite{qoin}, Qraft~\cite{qraft}, and Q-LEAR~\cite{qlear}, the majority of existing techniques cannot be applied to real noisy quantum computers and rely on statistical test oracles that are incompatible with modern error mitigation methods. Even the few approaches that support real hardware execution remain constrained by their dependence on full specifications and their limited applicability to optimization and variational circuits. 

The QOPS~\cite{qops} approach was evaluated against the existing testing methods using statistical test oracles and demonstrated superior performance. Consequently, in this study, we compare our results against QOPS~\cite{qops}, as it is the only existing approach that enables a direct and fair comparison with \method.

\section{Conclusion}\label{sec:conclusion}
QOPS~\cite{qops} was proposed to address key challenges in quantum software testing, including incompatible test case definitions, the high cost of full program specifications, and limited compatibility with error mitigation techniques. It defines test cases using Pauli strings, eliminating the need for explicit input states and enabling testing of search- and optimization-based circuits, while its novel oracle reduces reliance on complete specifications. However, QOPS relies on random test generation, which limits test budget efficiency and fault detection effectiveness. In this work, we extended the original QOPS approach as \method, a search-based variant that employed systematic test generation strategies to more effectively explore the search space. We also strengthened the empirical evaluation by scaling the analysis to larger quantum circuits of up to 29 qubits and by considering three different quantum computing architectures (IBM, IQM, and Quantinuum), both with and without error mitigation. Our experimental results demonstrated that \method substantially outperformed the original QOPS approach in detecting subtle faults with an average detection score of 100\%. Moreover, we also demonstrated that \method is architecturally portable and can be executed across multiple quantum computing platforms.

In this study, we also observed that a single quantum fault may exhibit a one-to-many relationship with distinct failing test cases, which complicates the localization process. We aim to further investigate the relationship between identified test cases and the underlying fault locations. To this end, in the future, we will study and adapt approaches such as spectrum-based fault localization~\cite{wong2016survey,sbflBook} and search-based localization methods~\cite{sbfl} to the quantum domain. Additionally, we plan to explore quantum-specific strategies, including circuit slicing~\cite{slicingtool} and lightweight tomography-inspired techniques such as classical shadows~\cite{clshadow}, to improve the precision and interpretability of fault localization in quantum circuits. For future work, we plan to explore this research direction. 

\section*{Acknowledgments}
Qu-Test project (Project \#299827) funded by the Research Council of Norway supports this work. S. Ali is also supported by Oslo Metropolitan University's Quantum Hub and Simula's internal strategic project on quantum software engineering. P. Arcaini is supported by the ASPIRE grant No. JPMJAP2301, JST. We acknowledge the use of IBM Quantum services for this work. The views expressed are those of the authors and do not reflect the official policy or position of IBM or the IBM Quantum team.



\begin{thebibliography}{10}
\providecommand{\url}[1]{#1}
\csname url@samestyle\endcsname
\providecommand{\newblock}{\relax}
\providecommand{\bibinfo}[2]{#2}
\providecommand{\BIBentrySTDinterwordspacing}{\spaceskip=0pt\relax}
\providecommand{\BIBentryALTinterwordstretchfactor}{4}
\providecommand{\BIBentryALTinterwordspacing}{\spaceskip=\fontdimen2\font plus
\BIBentryALTinterwordstretchfactor\fontdimen3\font minus \fontdimen4\font\relax}
\providecommand{\BIBforeignlanguage}[2]{{%
\expandafter\ifx\csname l@#1\endcsname\relax
\typeout{** WARNING: IEEEtran.bst: No hyphenation pattern has been}%
\typeout{** loaded for the language `#1'. Using the pattern for}%
\typeout{** the default language instead.}%
\else
\language=\csname l@#1\endcsname
\fi
#2}}
\providecommand{\BIBdecl}{\relax}
\BIBdecl

\bibitem{qseRoadmapTOSEM2025}
\BIBentryALTinterwordspacing
J.~M. Murillo, J.~Garcia-Alonso, E.~Moguel, J.~Barzen, F.~Leymann, S.~Ali, T.~Yue, P.~Arcaini, R.~P\'{e}rez-Castillo, I.~Garc\'{\i}a-Rodr\'{\i}guez~de Guzm\'{a}n, M.~Piattini, A.~Ruiz-Cort\'{e}s, A.~Brogi, J.~Zhao, A.~Miranskyy, and M.~Wimmer, ``Quantum software engineering: Roadmap and challenges ahead,'' \emph{ACM Trans. Softw. Eng. Methodol.}, vol.~34, no.~5, May 2025. [Online]. Available: \url{https://doi.org/10.1145/3712002}
\BIBentrySTDinterwordspacing

\bibitem{qstSurveyJSEP2023}
\BIBentryALTinterwordspacing
A.~García de~la Barrera, I.~García-Rodríguez~de Guzmán, M.~Polo, and M.~Piattini, ``Quantum software testing: State of the art,'' \emph{Journal of Software: Evolution and Process}, vol.~35, no.~4, p. e2419, 2023. [Online]. Available: \url{https://onlinelibrary.wiley.com/doi/abs/10.1002/smr.2419}
\BIBentrySTDinterwordspacing

\bibitem{quantumSTroadmapTOSEM25}
\BIBentryALTinterwordspacing
N.~C. Leite~Ramalho, H.~Amario~de Souza, and M.~Lordello~Chaim, ``Testing and debugging quantum programs: The road to 2030,'' \emph{ACM Trans. Softw. Eng. Methodol.}, vol.~34, no.~5, May 2025. [Online]. Available: \url{https://doi.org/10.1145/3715106}
\BIBentrySTDinterwordspacing

\bibitem{Basics}
N.~S. Yanofsky and M.~A. Mannucci, \emph{Quantum computing for computer scientists}.\hskip 1em plus 0.5em minus 0.4em\relax Cambridge University Press, 2008.

\bibitem{Combinatorial}
X.~Wang, P.~Arcaini, T.~Yue, and S.~Ali, ``Application of combinatorial testing to quantum programs,'' in \emph{2021 IEEE 21st International Conference on Software Quality, Reliability and Security (QRS)}, 2021, pp. 179--188.

\bibitem{qucatASE23tool}
\BIBentryALTinterwordspacing
------, ``{QuCAT}: A combinatorial testing tool for quantum software,'' in \emph{Proceedings of the 38th IEEE/ACM International Conference on Automated Software Engineering}, ser. ASE '23.\hskip 1em plus 0.5em minus 0.4em\relax IEEE Press, 2024, pp. 2066--2069. [Online]. Available: \url{https://doi.org/10.1109/ASE56229.2023.00062}
\BIBentrySTDinterwordspacing

\bibitem{fuzz}
J.~Wang, F.~Ma, and Y.~Jiang, ``{Poster: Fuzz Testing of Quantum Program},'' \emph{Proceedings - 2021 IEEE 14th International Conference on Software Testing, Verification and Validation, ICST 2021}, vol.~2, no.~1, pp. 466--469, 2021.

\bibitem{property}
\BIBentryALTinterwordspacing
S.~Honarvar, M.~R. Mousavi, and R.~Nagarajan, ``Property-based testing of quantum programs in {Q\#},'' in \emph{Proceedings of the IEEE/ACM 42nd International Conference on Software Engineering Workshops}, ser. ICSEW'20.\hskip 1em plus 0.5em minus 0.4em\relax New York, NY, USA: Association for Computing Machinery, 2020, pp. 430--435. [Online]. Available: \url{https://doi.org/10.1145/3387940.3391459}
\BIBentrySTDinterwordspacing

\bibitem{property2}
G.~Pontolillo and M.~R. Mousavi, ``A multi-lingual benchmark for property-based testing of quantum programs,'' in \emph{Proceedings of the 3rd International Workshop on Quantum Software Engineering}, ser. Q-SE '22.\hskip 1em plus 0.5em minus 0.4em\relax New York, NY, USA: Association for Computing Machinery, 2023, pp. 1--7.

\bibitem{GarciaICSOC2024}
\BIBentryALTinterwordspacing
A.~Garc\'{\i}a de~la Barrera, M.~A. Serrano, I.~Garc\'{\i}a-Rodr\'{\i}guez~de Guzm\'{a}n, M.~Polo, and M.~Piattini, ``Generating property-based tests for quantum algorithms,'' in \emph{Service-Oriented Computing – ICSOC 2024 Workshops: ASOCA, AI-PA, WESOACS, GAISS, LAIS, AI on Edge, RTSEMS, SQS, SOCAISA, SOC4AI and Satellite Events, Tunis, Tunisia, December 3–6, 2024, Revised Selected Papers, Part II}.\hskip 1em plus 0.5em minus 0.4em\relax Berlin, Heidelberg: Springer-Verlag, 2025, pp. 15--25. [Online]. Available: \url{https://doi.org/10.1007/978-981-96-7423-7_2}
\BIBentrySTDinterwordspacing

\bibitem{AbreuQSE22}
\BIBentryALTinterwordspacing
R.~Abreu, J.~a.~P. Fernandes, L.~Llana, and G.~Tavares, ``Metamorphic testing of oracle quantum programs,'' in \emph{Proceedings of the 3rd International Workshop on Quantum Software Engineering}, ser. Q-SE '22.\hskip 1em plus 0.5em minus 0.4em\relax New York, NY, USA: Association for Computing Machinery, 2023, pp. 16--23. [Online]. Available: \url{https://doi.org/10.1145/3528230.3529189}
\BIBentrySTDinterwordspacing

\bibitem{mutation}
\BIBentryALTinterwordspacing
E.~Mendiluze, S.~Ali, P.~Arcaini, and T.~Yue, ``Muskit: A mutation analysis tool for quantum software testing,'' in \emph{Proceedings of the 36th IEEE/ACM International Conference on Automated Software Engineering}, ser. ASE '21.\hskip 1em plus 0.5em minus 0.4em\relax IEEE Press, 2022, pp. 1266--1270. [Online]. Available: \url{https://doi.org/10.1109/ASE51524.2021.9678563}
\BIBentrySTDinterwordspacing

\bibitem{mutation2}
\BIBentryALTinterwordspacing
D.~Fortunato, J.~Campos, and R.~Abreu, ``{QMutPy}: a mutation testing tool for quantum algorithms and applications in {Qiskit},'' in \emph{Proceedings of the 31st ACM SIGSOFT International Symposium on Software Testing and Analysis}, ser. ISSTA 2022.\hskip 1em plus 0.5em minus 0.4em\relax New York, NY, USA: Association for Computing Machinery, 2022, pp. 797--800. [Online]. Available: \url{https://doi.org/10.1145/3533767.3543296}
\BIBentrySTDinterwordspacing

\bibitem{quantumMutantsEMSE2025}
\BIBentryALTinterwordspacing
E.~Mendiluze~Usandizaga, S.~Ali, T.~Yue, and P.~Arcaini, ``Quantum circuit mutants: {Empirical} analysis and recommendations,'' \emph{Empirical Software Engineering}, vol.~30, no.~3, pp. 1--35, Apr. 2025. [Online]. Available: \url{https://doi.org/10.1007/s10664-025-10643-z}
\BIBentrySTDinterwordspacing

\bibitem{FortunatoICSE22}
\BIBentryALTinterwordspacing
D.~Fortunato, J.~Campos, and R.~Abreu, ``Mutation testing of quantum programs written in {QISKit},'' in \emph{Proceedings of the ACM/IEEE 44th International Conference on Software Engineering: Companion Proceedings}, ser. ICSE '22.\hskip 1em plus 0.5em minus 0.4em\relax New York, NY, USA: Association for Computing Machinery, 2022, pp. 358--359. [Online]. Available: \url{https://doi.org/10.1145/3510454.3528649}
\BIBentrySTDinterwordspacing

\bibitem{qops}
\BIBentryALTinterwordspacing
A.~Muqeet, S.~Ali, and P.~Arcaini, ``Quantum program testing through commuting pauli strings on {IBM}'s quantum computers,'' in \emph{Proceedings of the 39th IEEE/ACM International Conference on Automated Software Engineering}, ser. ASE '24.\hskip 1em plus 0.5em minus 0.4em\relax New York, NY, USA: Association for Computing Machinery, 2024, pp. 2130--2141. [Online]. Available: \url{https://doi.org/10.1145/3691620.3695275}
\BIBentrySTDinterwordspacing

\bibitem{ssbse_survey}
\BIBentryALTinterwordspacing
M.~Harman, S.~A. Mansouri, and Y.~Zhang, ``Search-based software engineering: Trends, techniques and applications,'' \emph{ACM Comput. Surv.}, vol.~45, no.~1, Dec. 2012. [Online]. Available: \url{https://doi.org/10.1145/2379776.2379787}
\BIBentrySTDinterwordspacing

\bibitem{dirac}
P.~A.~M. Dirac, ``A new notation for quantum mechanics,'' in \emph{Mathematical Proceedings of the Cambridge Philosophical Society}, vol.~35, no.~3.\hskip 1em plus 0.5em minus 0.4em\relax Cambridge University Press, 1939, pp. 416--418.

\bibitem{basic}
D.~P. DiVincenzo, ``Quantum gates and circuits,'' \emph{Proceedings of the Royal Society of London. Series A: Mathematical, Physical and Engineering Sciences}, vol. 454, no. 1969, pp. 261--276, 1998.

\bibitem{fastpart}
\BIBentryALTinterwordspacing
B.~Reggio, N.~Butt, A.~Lytle, and P.~Draper, ``Fast partitioning of {Pauli} strings into commuting families for optimal expectation value measurements of dense operators,'' \emph{Phys. Rev. A}, vol. 110, p. 022606, Aug 2024. [Online]. Available: \url{https://link.aps.org/doi/10.1103/PhysRevA.110.022606}
\BIBentrySTDinterwordspacing

\bibitem{Paulistrings2}
\BIBentryALTinterwordspacing
T.~Kurita, M.~Morita, H.~Oshima, and S.~Sato, ``Pauli {String} {Partitioning} {Algorithm} with the {Ising} {Model} for {Simultaneous} {Measurements},'' \emph{The Journal of Physical Chemistry A}, vol. 127, no.~4, pp. 1068--1080, Feb. 2023. [Online]. Available: \url{https://doi.org/10.1021/acs.jpca.2c06453}
\BIBentrySTDinterwordspacing

\bibitem{noise_benchmark1}
\BIBentryALTinterwordspacing
S.~Resch and U.~R. Karpuzcu, ``{Benchmarking Quantum Computers and the Impact of Quantum Noise},'' \emph{ACM Comput. Surv.}, vol.~54, no.~7, jul 2021. [Online]. Available: \url{https://doi.org/10.1145/3464420}
\BIBentrySTDinterwordspacing

\bibitem{decoherence_def}
\BIBentryALTinterwordspacing
R.~Alicki, ``{Decoherence and the Appearance of a Classical World in Quantum Theory},'' \emph{Journal of Physics A: Mathematical and General}, vol.~37, no.~5, p. 1948, feb 2004. [Online]. Available: \url{https://dx.doi.org/10.1088/0305-4470/37/5/B02}
\BIBentrySTDinterwordspacing

\bibitem{crosstalkgatenoise}
\BIBentryALTinterwordspacing
T.~Ayral, F.-M.~L. R\'{e}gent, Z.~Saleem, Y.~Alexeev, and M.~Suchara, ``Quantum divide and compute: Exploring the effect of different noise sources,'' \emph{SN Comput. Sci.}, vol.~2, no.~3, Mar. 2021. [Online]. Available: \url{https://doi.org/10.1007/s42979-021-00508-9}
\BIBentrySTDinterwordspacing

\bibitem{simulators}
\BIBentryALTinterwordspacing
A.~Jamadagni, A.~M. L{\"{a}}uchli, and C.~Hempel, ``Benchmarking quantum computer simulation software packages,'' \emph{CoRR}, vol. abs/2401.09076, 2024. [Online]. Available: \url{https://doi.org/10.48550/arXiv.2401.09076}
\BIBentrySTDinterwordspacing

\bibitem{simulationlimit}
\BIBentryALTinterwordspacing
Y.~Zhou, E.~M. Stoudenmire, and X.~Waintal, ``{What Limits the Simulation of Quantum Computers?}'' \emph{Phys. Rev. X}, vol.~10, p. 041038, Nov 2020. [Online]. Available: \url{https://link.aps.org/doi/10.1103/PhysRevX.10.041038}
\BIBentrySTDinterwordspacing

\bibitem{quitoASE21tool}
\BIBentryALTinterwordspacing
X.~Wang, P.~Arcaini, T.~Yue, and S.~Ali, ``Quito: A coverage-guided test generator for quantum programs,'' in \emph{Proceedings of the 36th IEEE/ACM International Conference on Automated Software Engineering}, ser. ASE '21.\hskip 1em plus 0.5em minus 0.4em\relax IEEE Press, 2022, pp. 1237--1241. [Online]. Available: \url{https://doi.org/10.1109/ASE51524.2021.9678798}
\BIBentrySTDinterwordspacing

\bibitem{search}
\BIBentryALTinterwordspacing
------, ``{QuSBT}: Search-based testing of quantum programs,'' in \emph{Proceedings of the ACM/IEEE 44th International Conference on Software Engineering: Companion Proceedings}, ser. ICSE '22.\hskip 1em plus 0.5em minus 0.4em\relax New York, NY, USA: Association for Computing Machinery, 2022, pp. 173--177. [Online]. Available: \url{https://doi.org/10.1145/3510454.3516839}
\BIBentrySTDinterwordspacing

\bibitem{qlear}
\BIBentryALTinterwordspacing
A.~Muqeet, S.~Ali, T.~Yue, and P.~Arcaini, ``A machine learning-based error mitigation approach for reliable software development on {IBM}'s quantum computers,'' in \emph{Companion Proceedings of the 32nd ACM International Conference on the Foundations of Software Engineering}, ser. FSE 2024.\hskip 1em plus 0.5em minus 0.4em\relax New York, NY, USA: Association for Computing Machinery, 2024, pp. 80--91. [Online]. Available: \url{https://doi.org/10.1145/3663529.3663830}
\BIBentrySTDinterwordspacing

\bibitem{qoin}
A.~Muqeet, T.~Yue, S.~Ali, and P.~Arcaini, ``Mitigating noise in quantum software testing using machine learning,'' \emph{IEEE Transactions on Software Engineering}, vol.~50, no.~11, pp. 2947--2961, 2024.

\bibitem{Huang2019}
\BIBentryALTinterwordspacing
Y.~Huang and M.~Martonosi, ``Statistical assertions for validating patterns and finding bugs in quantum programs,'' in \emph{Proceedings of the 46th International Symposium on Computer Architecture}, ser. ISCA '19.\hskip 1em plus 0.5em minus 0.4em\relax New York, NY, USA: Association for Computing Machinery, 2019, pp. 541--553. [Online]. Available: \url{https://doi.org/10.1145/3307650.3322213}
\BIBentrySTDinterwordspacing

\bibitem{mqt}
N.~Quetschlich, L.~Burgholzer, and R.~Wille, ``{{MQT Bench}}: Benchmarking software and design automation tools for quantum computing,'' \emph{{Quantum}}, 2023, {{MQT Bench}} is available at \url{https://www.cda.cit.tum.de/mqtbench/}.

\bibitem{sourcecode}
A.~Muqeet, A.~Shaukat, and P.~Arcaini, ``{SB-QOPS} source code,'' \url{https://github.com/AsmarMuqeet/SB-QOPS}, 2026.

\bibitem{mealpy}
\BIBentryALTinterwordspacing
N.~Van~Thieu and S.~Mirjalili, ``{MEALPY}: An open-source library for latest meta-heuristic algorithms in {Python},'' \emph{J. Syst. Archit.}, vol. 139, no.~C, Jun. 2023. [Online]. Available: \url{https://doi.org/10.1016/j.sysarc.2023.102871}
\BIBentrySTDinterwordspacing

\bibitem{mealpyoptimizer}
N.~Van~Thieu, S.~D. Barma, T.~Van~Lam, O.~Kisi, and A.~Mahesha, ``Groundwater level modeling using augmented artificial ecosystem optimization,'' \emph{Journal of Hydrology}, vol. 617, p. 129034, 2023.

\bibitem{qiskit}
\BIBentryALTinterwordspacing
A.~Javadi-Abhari, M.~Treinish, K.~Krsulich, C.~J. Wood, J.~Lishman, J.~Gacon, S.~Martiel, P.~D. Nation, L.~S. Bishop, A.~W. Cross, B.~R. Johnson, and J.~M. Gambetta, ``Quantum computing with {Qiskit},'' \emph{CoRR}, vol. abs/2405.08810, 2024. [Online]. Available: \url{https://doi.org/10.48550/arXiv.2405.08810}
\BIBentrySTDinterwordspacing

\bibitem{IBMQuantum}
{IBM}, ``{IBM Quantum},'' \url{https://quantum.cloud.ibm.com/}, 2026, accessed: 2026-02-02.

\bibitem{GoogleQuantum}
{Google}, ``Google quantum,'' \url{https://quantumai.google/}, 2026, accessed: 2026-02-02.

\bibitem{IQMQuantum}
{IQM}, ``{IQM Quantum},'' \url{https://resonance.meetiqm.com/}, 2026, accessed: 2026-02-02.

\bibitem{QuantinuumQuantum}
{Quantinuum}, ``Quantinuum nexus,'' \url{https://nexus.quantinuum.com/}, 2026, accessed: 2026-02-02.

\bibitem{mitiq}
\BIBentryALTinterwordspacing
R.~LaRose, A.~Mari, S.~Kaiser, P.~J. Karalekas, A.~A. Alves, P.~Czarnik, M.~E. Mandouh, M.~H. Gordon, Y.~Hindy, A.~Robertson, P.~Thakre, M.~Wahl, D.~Samuel, R.~Mistri, M.~Tremblay, N.~Gardner, N.~T. Stemen, N.~Shammah, and W.~J. Zeng, ``Mitiq: A software package for error mitigation on noisy quantum computers,'' \emph{Quantum}, vol.~6, p. 774, Aug 2022. [Online]. Available: \url{https://doi.org/10.22331/q-2022-08-11-774}
\BIBentrySTDinterwordspacing

\bibitem{peccost}
\BIBentryALTinterwordspacing
Y.~Ma and M.~S. Kim, ``Limitations of probabilistic error cancellation for open dynamics beyond sampling overhead,'' \emph{Phys. Rev. A}, vol. 109, p. 012431, Jan 2024. [Online]. Available: \url{https://link.aps.org/doi/10.1103/PhysRevA.109.012431}
\BIBentrySTDinterwordspacing

\bibitem{recomended_effectsize}
E.~Tomczak and M.~Tomczak, ``The need to report effect size estimates revisited. an overview of some recommended measures of effect size,'' \emph{TRENDS in Sport Sciences}, vol.~21, no.~1, 2014.

\bibitem{statistics2}
\BIBentryALTinterwordspacing
A.~Arcuri and L.~Briand, ``{A Practical Guide for Using Statistical Tests to Assess Randomized Algorithms in Software Engineering},'' in \emph{Proceedings of the 33rd International Conference on Software Engineering}, ser. ICSE '11.\hskip 1em plus 0.5em minus 0.4em\relax New York, NY, USA: Association for Computing Machinery, 2011, pp. 1--10. [Online]. Available: \url{https://doi.org/10.1145/1985793.1985795}
\BIBentrySTDinterwordspacing

\bibitem{commonf3}
K.~P. Murphy, ``Performance evaluation of binary classifiers,'' \emph{Technical report: Technical Report}, 2007.

\bibitem{pea}
\BIBentryALTinterwordspacing
Y.~Kim, C.~J. Wood, T.~J. Yoder, S.~T. Merkel, J.~M. Gambetta, K.~Temme, and A.~Kandala, ``Scalable error mitigation for noisy quantum circuits produces competitive expectation values,'' \emph{Nature Physics}, vol.~19, no.~5, pp. 752--759, May 2023. [Online]. Available: \url{https://doi.org/10.1038/s41567-022-01914-3}
\BIBentrySTDinterwordspacing

\bibitem{threats}
D.~S. Cruzes and L.~ben Othmane, ``Threats to validity in empirical software security research,'' in \emph{Empirical research for software security}.\hskip 1em plus 0.5em minus 0.4em\relax CRC Press, 2017, pp. 275--300.

\bibitem{speed}
\BIBentryALTinterwordspacing
\emph{Classical and quantum computing}.\hskip 1em plus 0.5em minus 0.4em\relax New York, NY: Springer New York, 2007, pp. 203--217. [Online]. Available: \url{https://doi.org/10.1007/978-0-387-36944-0_13}
\BIBentrySTDinterwordspacing

\bibitem{ssbsetesting1}
P.~McMinn, ``Search-based software testing: Past, present and future,'' in \emph{2011 IEEE Fourth International Conference on Software Testing, Verification and Validation Workshops}, 2011, pp. 153--163.

\bibitem{harman2009theoretical}
M.~Harman and P.~McMinn, ``A theoretical and empirical study of search-based testing: Local, global, and hybrid search,'' \emph{IEEE Transactions on Software Engineering}, vol.~36, no.~2, pp. 226--247, 2009.

\bibitem{HarmanICST2015}
M.~Harman, Y.~Jia, and Y.~Zhang, ``Achievements, open problems and challenges for search based software testing,'' in \emph{2015 IEEE 8th International Conference on Software Testing, Verification and Validation (ICST)}, 2015, pp. 1--12.

\bibitem{coverage}
S.~Ali, P.~Arcaini, X.~Wang, and T.~Yue, ``Assessing the effectiveness of input and output coverage criteria for testing quantum programs,'' in \emph{2021 IEEE 14th International Conference on Software Testing, Validation and Verification (ICST)}, 2021, pp. 13--23.

\bibitem{LongTOSEM2024}
\BIBentryALTinterwordspacing
P.~Long and J.~Zhao, ``Testing multi-subroutine quantum programs: From unit testing to integration testing,'' \emph{ACM Trans. Softw. Eng. Methodol.}, vol.~33, no.~6, jun 2024. [Online]. Available: \url{https://doi.org/10.1145/3656339}
\BIBentrySTDinterwordspacing

\bibitem{bug1}
J.~Luo, P.~Zhao, Z.~Miao, S.~Lan, and J.~Zhao, ``A comprehensive study of bug fixes in quantum programs,'' in \emph{2022 IEEE International Conference on Software Analysis, Evolution and Reengineering (SANER)}, 2022, pp. 1239--1246.

\bibitem{bug2}
P.~Zhao, J.~Zhao, and L.~Ma, ``Identifying bug patterns in quantum programs,'' in \emph{2021 IEEE/ACM 2nd International Workshop on Quantum Software Engineering (Q-SE)}, 2021, pp. 16--21.

\bibitem{MiranskyyICSE2020}
\BIBentryALTinterwordspacing
A.~Miranskyy, L.~Zhang, and J.~Doliskani, ``Is your quantum program bug-free?'' in \emph{Proceedings of the ACM/IEEE 42nd International Conference on Software Engineering: New Ideas and Emerging Results}, ser. ICSE-NIER '20.\hskip 1em plus 0.5em minus 0.4em\relax New York, NY, USA: Association for Computing Machinery, 2020, pp. 29--32. [Online]. Available: \url{https://doi.org/10.1145/3377816.3381731}
\BIBentrySTDinterwordspacing

\bibitem{bug3}
P.~Zhao, J.~Zhao, Z.~Miao, and S.~Lan, ``{Bugs4Q}: A benchmark of real bugs for quantum programs,'' in \emph{2021 36th IEEE/ACM International Conference on Automated Software Engineering (ASE)}, 2021, pp. 1373--1376.

\bibitem{noise}
\BIBentryALTinterwordspacing
S.~Resch and U.~R. Karpuzcu, ``Benchmarking quantum computers and the impact of quantum noise,'' \emph{ACM Comput. Surv.}, vol.~54, no.~7, Jul. 2021. [Online]. Available: \url{https://doi.org/10.1145/3464420}
\BIBentrySTDinterwordspacing

\bibitem{noisesource}
\BIBentryALTinterwordspacing
K.~Georgopoulos, C.~Emary, and P.~Zuliani, ``Modeling and simulating the noisy behavior of near-term quantum computers,'' \emph{Phys. Rev. A}, vol. 104, p. 062432, Dec 2021. [Online]. Available: \url{https://link.aps.org/doi/10.1103/PhysRevA.104.062432}
\BIBentrySTDinterwordspacing

\bibitem{qem}
\BIBentryALTinterwordspacing
Z.~Cai, R.~Babbush, S.~C. Benjamin, S.~Endo, W.~J. Huggins, Y.~Li, J.~R. McClean, and T.~E. O'Brien, ``Quantum error mitigation,'' \emph{Rev. Mod. Phys.}, vol.~95, p. 045005, Dec 2023. [Online]. Available: \url{https://link.aps.org/doi/10.1103/RevModPhys.95.045005}
\BIBentrySTDinterwordspacing

\bibitem{qraft}
\BIBentryALTinterwordspacing
T.~Patel and D.~Tiwari, ``Qraft: reverse your quantum circuit and know the correct program output,'' in \emph{Proceedings of the 26th ACM International Conference on Architectural Support for Programming Languages and Operating Systems}, ser. ASPLOS '21.\hskip 1em plus 0.5em minus 0.4em\relax New York, NY, USA: Association for Computing Machinery, 2021, pp. 443--455. [Online]. Available: \url{https://doi.org/10.1145/3445814.3446743}
\BIBentrySTDinterwordspacing

\bibitem{wong2016survey}
\BIBentryALTinterwordspacing
W.~E. Wong, R.~Gao, Y.~Li, R.~Abreu, and F.~Wotawa, ``A survey on software fault localization,'' \emph{IEEE Trans. Softw. Eng.}, vol.~42, no.~8, pp. 707--740, aug 2016. [Online]. Available: \url{https://doi.org/10.1109/TSE.2016.2521368}
\BIBentrySTDinterwordspacing

\bibitem{sbflBook}
\BIBentryALTinterwordspacing
X.~Xie and B.~Xu, \emph{Essential Spectrum-based Fault Localization}.\hskip 1em plus 0.5em minus 0.4em\relax Springer, 2021. [Online]. Available: \url{https://doi.org/10.1007/978-981-33-6179-9}
\BIBentrySTDinterwordspacing

\bibitem{sbfl}
\BIBentryALTinterwordspacing
C.~Liu, Y.~Lei, H.~Xie, J.~Wang, Y.~Yu, and D.~Lo, ``Survey on learning-based dynamic fault localization: From traditional machine learning to large language models,'' \emph{ACM Comput. Surv.}, Jan. 2026, just Accepted. [Online]. Available: \url{https://doi.org/10.1145/3787202}
\BIBentrySTDinterwordspacing

\bibitem{slicingtool}
S.~A. Metwalli and R.~Van~Meter, ``A tool for debugging quantum circuits,'' in \emph{2022 IEEE International Conference on Quantum Computing and Engineering (QCE)}, 2022, pp. 624--634.

\bibitem{clshadow}
\BIBentryALTinterwordspacing
D.~E. Koh and S.~Grewal, ``Classical {S}hadows {W}ith {N}oise,'' \emph{{Quantum}}, vol.~6, p. 776, Aug. 2022. [Online]. Available: \url{https://doi.org/10.22331/q-2022-08-16-776}
\BIBentrySTDinterwordspacing

\end{thebibliography}
\end{document}